\documentclass[12pt,letterpaper]{article}

\usepackage{etoolbox}
\newtoggle{SUPPLEMENTAL}\toggletrue{SUPPLEMENTAL}
\togglefalse{SUPPLEMENTAL} %COMMENT OUT FOR SUPPLEMENTAL APPENDIX
\newcommand{\Supplemental}[2]{\iftoggle{SUPPLEMENTAL}{#1}{#2}}
\newtoggle{BLINDED}\toggletrue{BLINDED}
\togglefalse{BLINDED} % <-------- COMMENT OUT FOR BLINDED VERSION
\newcommand{\Blinded}[2]{\iftoggle{BLINDED}{#1}{#2}}

\usepackage{graphicx,grffile}
\usepackage{amsmath,amssymb,amsthm}
\usepackage{mathtools,dsfont,centernot}
\usepackage{caption}
\usepackage{booktabs,tabularx,threeparttable,longtable,ragged2e}
\usepackage{siunitx}
\usepackage[shortlabels]{enumitem}
\usepackage{alltt}
\usepackage[longnamesfirst]{natbib}
\usepackage{hypernat}
\usepackage[nodisplayskipstretch]{setspace}
\usepackage[bottom]{footmisc}
\usepackage{xr} %xr,xr-hyper}
\usepackage[final]{listings}
\lstdefinestyle{inlineR}{language=R,frame=none,basicstyle=\ttfamily,keywordstyle=\ttfamily,stringstyle=\ttfamily,keepspaces=true,showspaces=false,showstringspaces=false,breaklines=true,upquote=true,print,columns=fullflexible}
\newcommand{\code}{\lstinline}

\usepackage[hyphens]{url}
\usepackage{hyperref}

\usepackage[margin=1in,letterpaper]{geometry}
\usepackage[capitalize,noabbrev]{cleveref}

  \renewenvironment{thebibliography}[1]%
  {\begin{oldthebibliography}{#1}\setlength{\parskip}{0ex}\setlength{\itemsep}{0ex}}%
  {\end{oldthebibliography}}
\appto\TPTnoteSettings{\linespread{1}\footnotesize}

\DeclareGraphicsExtensions{.pdf,.png}
\graphicspath{ {../figures/} {./figures/} }

\newcommand{\numnornd}[1]{\num[round-mode=none,group-digits=integer]{#1}} % no rounding, but group-separator;  ,group-separator={\,}
\newcommand{\USD}[1]{\SI[round-precision=0,round-mode=places]{#1}[\$]{}}

\newcommand{\citeposs}[1]{\citeauthor{#1}'s (\citeyear{#1})}

\crefname{conjecture}{Conjecture}{Conjectures}
\crefname{section}{Section}{Sections}
\crefname{subsection}{Section}{Sections}
\crefname{subsubsection}{Section}{Sections}
\Crefname{conjecture}{Conjecture}{Conjectures}
\Crefname{section}{Section}{Sections}
\Crefname{subsection}{Section}{Sections}
\Crefname{subsubsection}{Section}{Sections}
\crefname{appendix}{Appendix}{Appendices}
\crefname{subappendix}{Appendix}{Appendices}
\crefname{subsubappendix}{Appendix}{Appendices}
\Crefname{appendix}{Appendix}{Appendices}
\Crefname{subappendix}{Appendix}{Appendices}
\Crefname{subsubappendix}{Appendix}{Appendices}
\crefname{equation}{}{}
\Crefname{equation}{Equation}{Equations}
\crefname{enumi}{}{}
\Crefname{enumi}{}{}

\crefname{assumption}{}{}
\Crefname{assumption}{Assumption}{Assumptions}
\Crefname{method}{Method}{Methods}
\newlist{steps}{enumerate}{1}
\setlist[steps]{label=\arabic*., ref=\arabic*, itemsep=0pt}
\crefname{stepsi}{Step}{Steps}
\Crefname{stepsi}{Step}{Steps}

\theoremstyle{plain}
\newtheorem{theorem}{Theorem}%[section]
\newtheorem{lemma}[theorem]{Lemma}
\newtheorem{proposition}[theorem]{Proposition}
\newtheorem{corollary}[theorem]{Corollary}
\newtheorem{method}{Method}

\theoremstyle{definition}

\newtheorem{assumption}{Assumption}

\usepackage{bm} %better bold math for vecf and matf

\newcommand{\SD}[1]{\mathrel{\mathrm{SD}_{#1}}}
\newcommand{\nonSD}[1]{\mathrel{\mathit{non}\mathrm{SD}_{#1}}}
\newcommand{\SDF}{\SD{\mathcal{F}}}
\newcommand{\nonSDF}{\nonSD{\mathcal{F}}}

\newcommand{\pconv}{\xrightarrow{p}}
\newcommand{\dconv}{\xrightarrow{d}}

\DeclareMathOperator{\Var}{Var}

\newcommand{\R}{{\mathbb R}}

\DeclareMathOperator{\E}{E} %{\mathbb{E}}
\let\Pr\relax \DeclareMathOperator{\Pr}{P} %comment out if Pr desired
\newcommand{\Pn}{{\mathbb P}_n}
\newcommand{\Pna}{{\mathbb P}_n^a}
\newcommand{\Pnb}{{\mathbb P}_n^b}
\newcommand{\PnaBS}{\tilde{\mathbb P}_n^a}
\newcommand{\PnbBS}{\tilde{\mathbb P}_n^b}

\newcommand{\Ga}{{\mathbb G}^a}
\newcommand{\Gb}{{\mathbb G}^b}
\newcommand{\Gd}{{\mathbb G}^\Delta}
\newcommand{\Gn}{{\mathbb G}_n}
\newcommand{\Gna}{{\mathbb G}_n^a}
\newcommand{\Gnb}{{\mathbb G}_n^b}
\newcommand{\Gnd}{{\mathbb G}_n^\Delta}
\newcommand{\GndBS}{\tilde{\mathbb G}_n^\Delta}

\newcommand{\limn}{\lim_{n\to\infty}}

\newcommand{\FWER}{\textrm{FWER}}
\newcommand{\NormDist}{\mathrm{N}}

\newcommand{\weaklyto}{\rightsquigarrow}
\newcommand{\diff}{\mathop{}\!\mathrm{d}} %upright egreg version: see https://tex.stackexchange.com/a/84308
\newcommand{\pD}[2]{\frac{\partial #1}{\partial #2}}

\newcommand{\independenT}[2]{\mathrel{\rlap{$#1#2$}\mkern2mu{#1#2}}}
\newcommand\independent{\protect\mathpalette{\protect\independenT}{\perp}} 
\providecommand{\abs}[1]{\lvert#1\rvert}
\providecommand{\norm}[1]{\lVert#1\rVert}

\let\originalleft\left
\let\originalright\right
\renewcommand{\left}{\mathopen{}\mathclose\bgroup\originalleft}
\renewcommand{\right}{\aftergroup\egroup\originalright}

\newcommand{\mockalph}[1]{}  % for BibTeX sorting
\allowdisplaybreaks[3]

\iftoggle{BLINDED}{}{
 \hypersetup{
  pdfauthor = {David M.\ Kaplan},
  pdfkeywords = {economics, econometrics, statistics},
  pdftitle = {Kaplan: inference on consensus ranking of distributions},
  pdfsubject = {econometrics},
  pdfpagemode = UseNone
 }
}

\title{Inference on Consensus Ranking of Distributions}

\author{\Blinded{[BLINDED]}{%
David M.\ Kaplan%
\thanks{Department of Economics, University of Missouri, \texttt{kaplandm@missouri.edu}}}}

\date{August 10, 2023} % \date{\today}

\begin{document}

\maketitle

\doublespacing
\begin{abstract}
Instead of testing for unanimous agreement, I propose learning how broad of a consensus favors one distribution over another (of earnings, productivity, asset returns, test scores, etc.).
Specifically, given a sample from each of two distributions, I propose statistical inference methods to learn about the set of utility functions for which the first distribution has higher expected utility than the second distribution.
With high probability, an ``inner'' confidence set is contained within this true set, while an ``outer'' confidence set contains the true set.
Such confidence sets can be formed by inverting a proposed multiple testing procedure that controls the familywise error rate.
Theoretical justification comes from empirical process results, given that very large classes of utility functions are generally Donsker (subject to finite moments).
The theory additionally justifies a uniform (over utility functions) confidence band of expected utility differences, as well as tests with a utility-based ``restricted stochastic dominance'' as either the null or alternative hypothesis.
Simulated and empirical examples illustrate the methodology.

\textit{JEL classification}: 
% https://www.aeaweb.org/jel/guide/jel.php
C29 % Single Equation Models; Single Variables: Other

\textit{Keywords}: 
confidence set, expected utility, multiple testing, stochastic dominance
\end{abstract}

\newcommand{\paperspacing}{\doublespacing}

\paperspacing

\Supplemental{\newpage}{}

\section{Introduction}

Different individuals with different preferences can disagree about which of two distributions is better, so economists have considered ways to capture consensus.
Within the standard expected utility framework, variations of stochastic dominance characterize when certain groups of people all agree which distribution is better, i.e., when expected utility is higher for one distribution given any utility function in a particular set.
For example, first-order stochastic dominance requires higher expected utility for any (non-decreasing) utility function, whereas second-order stochastic dominance relaxes the requirement to only concave (risk-averse) utility functions.

This setup can be interpreted broadly.
``Individuals'' can be replaced by other agents or organizations (or ``utility function'' replaced by social welfare function), and the ``distributions'' could be of earnings, asset returns, productivity, scores, or other observable measures.
Further, these could be conditional distributions, given partial information; for example, a firm could assess distributions only for firms with similar characteristics.
The overall goal may be description (e.g., differences across countries in observational data) or causal inference and policy decision-making (e.g., comparing randomized treatment and control groups).
For social welfare comparisons, this approach follows the idea of \citet{Vickrey1945} and \citet{Harsanyi1953} (later popularized by \citet{Rawls1971} as the ``veil of ignorance'' or ``original position'') to consider the expected utility of randomly becoming a member of one society versus another (see \cref{ft:OP} for quotations).

For both statistical and economic reasons, I propose methods to learn about \emph{the set of utility functions for which one distribution is preferred} to another.
Statistically, this set of utility functions provides more information than a single all-or-nothing hypothesis test of unanimous agreement (stochastic dominance).
Even without unanimous agreement, it is helpful to know whether there is a broad consensus over a large set of realistic utility functions.
Further, compared to testing a null hypothesis of stochastic dominance, my approach provides stronger evidence for a distribution being ``better.''
Incorrectly failing to reject a false null of dominance is a type II error, whose rate is not controlled.
In contrast, I control the error probability of making even one incorrect claim that a utility function is in the true set (details below).

Economically, without unanimous agreement, the expected utility perspective differs from the cumulative distribution function (CDF) perspective.
With unanimous agreement in the sense of first-order stochastic dominance, there are three equivalent characterizations: higher expected utility for any utility function; lower CDF; and higher quantile function.
The stochastic dominance testing literature has leveraged this to construct CDF-based tests whose results can be interpreted in terms of expected utility.
However, without unanimity, the equivalence disappears, so distinguishing the economic interpretations becomes important.
My consensus set of utility functions has a direct economic interpretation, showing which types of individuals would prefer a particular distribution over another.
These utility functions can be examined in light of estimates of utility function parameters like risk aversion.
Alternatively, given the quantile utility maximization framework \citep{Manski1988,Rostek2010}, the set of quantile indices at which one distribution has larger quantiles can be interpreted similarly, as the set of ``types'' of agents who would prefer that distribution.
From the CDF perspective, for income or consumption distributions, the set of values at which one CDF is below the other can be interpreted as the set of possible poverty lines for which one distribution has a lower poverty rate; this is \citeposs{Atkinson1987} ``restricted stochastic dominance'' (Condition I, p.\ 751), and ``poverty'' can be generalized to other types of distributions.
These are all economically meaningful, yet all different, so it seems valuable to develop methodology supporting each perspective.

To quantify uncertainty about this set of utility functions for which one distribution is preferred, I propose ``inner'' and ``outer'' confidence sets.
These quantify the statistical uncertainty due to having only a random sample from each distribution.
As with the set characterized by (1) of \citet{ArmstrongShen2015}, the inner confidence set is contained within the true consensus set with high probability, while the outer confidence set contains the true set with high probability (similar to a confidence set for an identified set).
The inner confidence set is thus conservative in the sense that it only contains utility functions from the true set with high probability, but it readily omits additional utility functions (from the true set) if there is too much uncertainty.
The true set is ``between'' the inner and outer confidence sets with high probability.

These confidence sets can be constructed by inverting a multiple testing procedure.
This link between inner confidence sets and multiple testing has been described in Section 2 of \citet{ArmstrongShen2015}.
The inner confidence set contains all utility functions for which lower expected utility is rejected in favor of higher expected utility while controlling the familywise error rate.
The probability of any false rejection is the probability of incorrectly including a utility function in the confidence set, so familywise error rate control implies correct coverage probability.
The outer confidence set contains all utility functions for which higher expected utility is not rejected, using the same logic as \citeposs{RomanoShaikh2010} confidence set for the identified set (e.g., see Lemma 2.1, pp.\ 172--173).
Alternatively, if the first distribution were known to yield higher expected utility for a single unknown utility function, then the utility function is partially identified, and the outer confidence set is indeed a confidence set for the identified set.
To choose among the many possible valid multiple testing procedures (and thus confidence sets), I use the same pointwise asymptotic size at each point.

To justify the multiple testing procedures and confidence sets, I use empirical process theory for the sample expected utility difference, indexed by utility functions.
Because utility functions are non-decreasing, this process generally has a Gaussian limit, subject to certain finite moments.
Further, this limit is consistently estimated by exchangeable bootstrap.

These theoretical results in turn justify other statistical methods.
Uniform (over utility functions) confidence bands for the expected utility difference can be constructed; these are more informative than the confidence sets, but more difficult to communicate and comprehend.
I also describe hypothesis tests for having higher expected utility for all utility functions in a specified set; the null can be either dominance or non-dominance, analogous to the CDF-based tests of non-dominance of \citet{DavidsonDuclos2013}.

There is a high-level parallel between my approach and that of \citet{GoldmanKaplan2018c} and \citet{distcomp}, but I make three main contributions beyond that previous work.
For reference, Method 5 of \citet{GoldmanKaplan2018c} is a multiple testing procedure for whether one CDF lies below another CDF at each possible point, i.e., it tests $H_{0r}\colon F_a(r)\le F_b(r)$ for all $r$ while controlling the (finite-sample) familywise error rate.
The first new contribution is my interpretation in terms of consensus.
As described above (and below), this new consensus framework can be applied to a variety of preference frameworks, including an application to the CDF setting of \citet{GoldmanKaplan2018c} through the interpretation of \citet{Atkinson1987} in terms of ``poverty.''
However, the original work of \citet{GoldmanKaplan2018c} said nothing of consensus or preferences or poverty, instead asking only the statistical question (p.\ 160), ``At which quantiles or CDF values do two distributions differ?â€™â€™
Second, I consider expected utility instead of CDF differences.
The CDF approach has natural advantages when data are top-coded or suffer from measurement error in the tails, but expected utility is a more conventional framework for preferences in economics that better connects to the economics literature.
Statistically, the CDF approach has the advantage of finite-sample inference based on order statistics when sampling is iid, but my empirical process-based inference is more readily extended to other types of sampling or preference frameworks.
Third, my object of interest is the true set of preference types (utility functions) for which one distribution is preferred over another, and I construct confidence sets for this object of interest.
As with the consensus framework, this confidence set interpretation can be applied back to the CDF setting (with ``preference types'' being poverty lines), but it is not found in the work of \citet{GoldmanKaplan2018c}.

My approach to learn about the consensus set can be applied to other distributional comparisons, too.
For example, there are many proposed metrics to capture ``inequality'' within a distribution, but often they depend on a user-chosen parameter.
Instead of picking one value (or several), we could learn about the set of parameter values for which one distribution is ``better'' (lower inequality), like such a set of $\epsilon$ in the \citet{Atkinson1970} inequality index, or $\alpha$ in the \citet[eqn.\ (22) and \S4.3]{CowellFlachaire2017} inequality index for ordinal variables.
That is, we can try to learn about the set of user-chosen parameter values within which there is a consensus about one distribution being better.
In the context of forecast comparison, we could learn about the set of utility functions for which one forecasting model has higher out-of-sample expected utility than another, building on the statistical framework of \citet{BarendsePatton2022}, who instead test the single null hypothesis of equal (or superior) predictive ability over a continuum of loss functions.
Other potential research questions are mentioned in the conclusion.

As already cited, \citet{ArmstrongShen2015} suggest a similar approach to a different economic problem in which the object of interest is the set of $x$ values at which the conditional average treatment effect is positive.
Although they do not write out the phrase ``inner confidence set,'' the characterization is the same as in their (1), and their two-sided extension in Appendix A has a pair of confidence sets with superscripts ``inner'' and ``outer.''
They also form their confidence sets by inverting multiple testing procedures, with critical values based on sup-$t$ statistic distributions, as in their (2).
Because the application differs, so do the lower-level details; for continuous $x$, they consider nonparametric kernel regression estimates of the conditional average treatment effect, whereas I consider expected utility differences over a continuum of utility functions.

\paragraph{Paper structure}
\Cref{sec:motivation} introduces the setting for the empirical illustrations of \cref{sec:emp}.
\Cref{sec:CS} describes the inner and outer confidence sets.
\Cref{sec:theory} formally develops methodology for confidence sets, a multiple testing procedure, and a uniform confidence band, along with theoretical results, illustrated by simulation in \cref{sec:sim}.
% 
% The supplementary appendix has additional (more technical) theoretical results, additional methodology, proofs not found in the main text, and an example algorithm to compute bootstrap critical values.
\Cref{sec:app-theory} contains additional (more technical) theoretical results.
\Cref{sec:additional} describes additional methodology: a stepdown multiple testing procedure and hypothesis tests for restricted stochastic dominance.
\Cref{sec:app-pfs} contains proofs not found in the main text.
\Cref{sec:BScv} provides an example algorithm to compute bootstrap critical values.

\paragraph{Notation}
Uppercase and lowercase indicate random and non-random variables, respectively, like random $Y$ with realization $y$.
Calligraphic letters indicate sets, such as $\mathcal{U}$ and $\mathcal{C}$.
Acronyms used include those for confidence set (CS), constant relative risk aversion (CRRA), coverage probability (CP), cumulative distribution function (CDF), data-generating process (DGP), empirical CDF (ECDF), familywise error rate (FWER), multiple testing procedure (MTP), and Vapnik--Chervonenkis (VC).

\section{Empirical Motivation}
\label{sec:motivation}

To fix ideas and motivate the new methodology, I introduce the empirical setting used later in \cref{sec:emp}.

The goal is to compare two distributions represented respectively by random variables $Y^a$ and $Y^b$, based on a random sample from each distribution.
Some examples are studied in \cref{sec:emp}.
For example, in the context of a randomized job training program, these could be earnings distributions for the population offered the program ($Y^a$) and the population not offered the program ($Y^b$).
Another example compares union and non-union wage distributions.
In each case, the goal is to summarize the statistical evidence in favor of the first distribution being ``better'' than the second distribution, using the economic framework of expected utility.

In the \cref{sec:emp} examples, the utility functions considered are shifted CRRA, where $\theta$ is the usual risk aversion parameter, and $s$ is the size of the shift (which can be interpreted as a subsistence level).
This shifted CRRA family can also be seen as a special case of the hyperbolic absolute risk aversion family of utility functions, restricted to those with decreasing absolute risk aversion; see (43) and Table I of \citet[p.\ 389]{Merton1971}, noting my $1-\theta$ is his $\gamma$, so my $\theta\ge0$ is his $\gamma\le1$.
% equiv w/ special case a=1-gamma of HARA utility: take U(W) from https://en.wikipedia.org/wiki/Hyperbolic_absolute_risk_aversion#Definition  and sub in W=y, gamma=1-theta, b=-s; then divide by theta and subtract 1/(1-theta).  (location/scale changes by constant do not affect behavior/decisions.)
Here, given $\theta$ and $s$,
\begin{equation}
\label{eqn:emp-CRRA}
u_{\theta,s}(y) =
\left\{
\begin{matrix}
\ln(y-s) & \textrm{if }\theta=1, \\
\frac{(y-s)^{1-\theta}-1}{1-\theta} & \textrm{if }\theta\ne1.
\end{matrix}
\right.
\end{equation}
The universe of utility functions analyzed is then
\begin{equation}
\label{eqn:emp-universe}
\mathcal{U} = \{u_{\theta,s} : 0\le\theta\le\bar{\theta} ,\; \underbar{s}\le s\le\bar{s} \} ,
\end{equation}
where the range of shifts $[\underbar{s},\bar{s}]$ depends on the example, as does upper bound $\bar{\theta}$.
More generally, $\mathcal{U}$ can collect utility functions with three or more parameters, or it can be the union of multiple sets; see \cref{sec:Donsker} for technical details.

The object of interest is the consensus set that contains all utility functions in the universe $\mathcal{U}$ for which $Y^a$ is preferred over $Y^b$, i.e., for which $Y^a$ has higher expected utility.
Mathematically, the consensus set is
\begin{equation}
\label{eqn:emp-consensus-set}
\mathcal{C} = \{ u\in\mathcal{U} : \E[u(Y^a)]>\E[u(Y^b)] \}.
\end{equation}
Given the shifted CRRA form in \cref{eqn:emp-CRRA,eqn:emp-universe}, the consensus set $\mathcal{C}$ in \cref{eqn:emp-consensus-set} can equivalently be expressed in terms of the risk aversion $\theta$ and shift $s$,
\begin{equation}
\label{eqn:C-tilde}
\mathcal{C} = \{ u_{\theta,s} : (\theta,s)\in\tilde{\mathcal{C}} \}
,\quad
\tilde{\mathcal{C}} = \{ (\theta,s) : \E[u_{\theta,s}(Y^a)]>\E[u_{\theta,s}(Y^b)],\; 0\le\theta\le\bar{\theta} ,\; \underbar{s}\le s\le\bar{s} \}
.
\end{equation}
This $\tilde{\mathcal{C}}$ can be visualized on a graph with horizontal axis $\theta$ and vertical axis $s$.

To learn about the consensus set $\mathcal{C}$ from data, we can follow the usual approach of computing a point estimate and confidence bounds.
The point estimate simply uses \cref{eqn:emp-consensus-set} but with sample averages $\hat{\E}(\cdot)$ instead of population expectations $\E(\cdot)$:
\begin{equation}
\label{eqn:consensus-set-est}
\hat{\mathcal{C}} = \{ u\in\mathcal{U} : \hat{\E}[u(Y^a)]>\hat{\E}[u(Y^b)] \}.
\end{equation}
To account for statistical uncertainty, I propose inner and outer confidence sets in \cref{sec:CS}.
Given shifted CRRA utility, these can be graphed in terms of $(\theta,s)$, as shown in \cref{sec:emp}: with high asymptotic probability, the true consensus set contains the inner confidence set and is contained by the outer confidence set.

\section{Confidence Sets and Multiple Testing}
\label{sec:CS}

For inference on the consensus set $\mathcal{C}$ in \cref{eqn:emp-consensus-set}, I propose the inner and outer confidence sets (CSs) described in this section.
\Cref{sec:CS-prop} describes properties and interpretation.
\Cref{sec:CS-comp} discusses how to construct the CSs.
\Cref{sec:CS-justification} discusses intuition for the formal justification, including some practical discussion of assumptions.

\subsection{Properties and Interpretation}
\label{sec:CS-prop}

The inner CS $\hat{\mathcal{C}}_1$ is contained within the true $\mathcal{C}$ with high asymptotic probability.
For confidence level $1-\alpha$,
\begin{equation}
\label{eqn:inner-CS-asyCP}
\lim_{n\to\infty} \Pr(\hat{\mathcal{C}}_1 \subseteq \mathcal{C}) \ge 1-\alpha .
\end{equation}
Thus, the inner CS provides a conservative assessment (a lower bound) of the consensus that prefers $Y^a$ over $Y^b$.
That is, we can feel reasonably sure that all utility functions in the inner CS are indeed part of the true consensus preferring $Y^a$, and there are actually even more elements of the true consensus set, but they just lack strong enough evidence to be included in the inner CS.
The inner CS is useful when facing skeptical belief that $Y^a\stackrel{d}{=}Y^b$ because it provides strong evidence that $Y^a$ is better for at least all the utility functions in $\hat{\mathcal{C}}_1$.

Alternatively, the outer CS $\hat{\mathcal{C}}_2$ contains the true consensus set $\mathcal{C}$ with high asymptotic probability:
\begin{equation}
\label{eqn:outer-CS-asyCP}
\lim_{n\to\infty} \Pr(\hat{\mathcal{C}}_2 \supseteq \mathcal{C}) \ge 1-\alpha .
\end{equation}
This essentially provides an ``upper bound'' for the consensus preferring $Y^a$ over $Y^b$.
Complementing the inner CS, the outer CS helps us learn how large the true consensus might be, but it may contain many utility functions that are not actually part of $\mathcal{C}$.
This can be useful in supporting $Y^a$ being better if $\hat{\mathcal{C}}_2=\mathcal{U}$ (the universe of utility functions considered), meaning it is possible that everyone (with $u\in\mathcal{U}$) prefers $Y^a$.
It is also useful if there is a common belief that $Y^a$ is unequivocally better than $Y^b$, but the outer CS is substantially smaller than $\mathcal{U}$, meaning that (contrary to common belief) the consensus supporting $Y^a$ is limited to something within $\hat{\mathcal{C}}_2$.

In terms of the consensus that prefers $Y^b$ over $Y^a$ (opposite above), the inner CS is the complement of the above outer CS $\hat{\mathcal{C}}_2$, and the outer CS is the complement of the above inner CS $\hat{\mathcal{C}}_1$.
That is, the same two sets can be interpreted in terms of either preference direction.

\subsection{Computation}
\label{sec:CS-comp}

Both types of confidence set can be constructed using results from a multiple testing procedure (MTP).
Consider the following families of null hypotheses indexed by utility function:
\begin{equation}
\label{eqn:H0}
\begin{split}
H_{0u}^{\le} \colon \E[u(Y^a)] \le \E[u(Y^b)]
,\quad u\in\mathcal{U} , \\
H_{0u}^{\ge} \colon \E[u(Y^a)] \ge \E[u(Y^b)]
,\quad u\in\mathcal{U} .
\end{split}
\end{equation}
The inner CS collects all utility functions for which $H_{0u}^{\le}$ is rejected in favor of the alternative $\E[u(Y^a)]>\E[u(Y^b)]$, which is equivalent to $u\in\mathcal{C}$.
The outer CS collects all utility functions for which $H_{0u}^{\ge}$ is not rejected, meaning $u\in\mathcal{C}$ is not rejected.
Mathematically,
\begin{equation}
\label{eqn:CS-based-on-MTP}
\hat{\mathcal{C}}_1 = \{ u\in\mathcal{U} : H_{0u}^{\le}\textrm{ rejected by MTP} \}
,\quad
\hat{\mathcal{C}}_2 = \{ u\in\mathcal{U} : H_{0u}^{\ge}\textrm{ not rejected by MTP} \} .
\end{equation}
Intuitively, this fits the interpretation of \cref{sec:CS-prop}: strong evidence is required in order to reject $H_{0u}^{\le}$ and include a utility function in the inner CS, whereas utility functions are included in the outer CS unless there is strong enough evidence to reject $H_{0u}^{\ge}$ and exclude them.
If there is a lot of statistical uncertainty, then an MTP tends not to reject, and thus the inner CS tends to be small and the outer CS tends to be large, analogous to a wide confidence interval (small lower endpoint, large upper endpoint).

In turn, a basic MTP can be computed using $t$-statistics and a bootstrap critical value that accounts for multiple testing.
Details of the following steps are in and implemented in the provided code.
Consider $H_{0u}^{\le}$; the $H_{0u}^{\ge}$ case is equivalent after switching $Y^a$ and $Y^b$.
First, compute a conventional $t$-statistic for $H_{0u}^{\le}$ for each $u\in\mathcal{U}$.
In practice, a fine grid over $\mathcal{U}$ is used; in the shifted CRRA example, this is equivalent to a grid for $(\theta,s)$.
Second, compute the bootstrap critical value.
A valid but conservative approach is to compute bootstrap-world $t$-statistics, but centered at $\hat{\E}[u(Y^a)]-\hat{\E}[u(Y^b)]$ instead of centered at zero.
The highest (over $u\in\mathcal{U}$) such ``$t$-statistic'' is computed for each bootstrap sample, and then the $1-\alpha$ quantile is taken across samples.
Third, reject any $H_{0u}^{\le}$ for which the $t$-statistic exceeds the critical value, and inclue that $u$ in the inner CS.

Power can be improved (while still controlling FWER) by the stepdown procedure in \cref{meth:stepdown}, or by pre-testing.

\subsection{Justification}
\label{sec:CS-justification}

While complete formal statements are left to \cref{sec:theory}, some intuition for the theoretical justification is given here.
The focus is on the inner CS.

First, the inner CS's asymptotic coverage is determined by the underlying MTP's familywise error rate (FWER), which is the probability of making even just one false rejection.
From \cref{eqn:H0,eqn:CS-based-on-MTP}, $\hat{\mathcal{C}}_1 \subseteq \mathcal{C}$ if and only if there are zero false rejections.
Thus, if asymptotically the FWER is below $\alpha$, then the coverage probability is at least $1-\alpha$:
\begin{equation}
\label{eqn:FWER-CP}
 \Pr(\hat{\mathcal{C}}_1\subseteq\mathcal{C})
=
 \Pr(\textrm{no true }H_{0u}^{\le}\textrm{ rejected})
= 1 - \overbrace{\Pr(\textrm{reject any true }H_{0u}^{\le})}^{\equiv\FWER\le\alpha+o(1)}
\ge 1-\alpha+o(1) .
\end{equation}

Second, the MTP's validity in turn depends on bootstrap consistency for the empirical process.
(Further taking the supremum of the process is simply the continuous mapping theorem.)
Given iid sampling, exchangeable bootstraps (including the usual empirical bootstrap as a special case) are valid under essentially the same conditions required for the empirical process to have a Gaussian limit, so besides iid sampling, the key assumption is that the utility function universe $\mathcal{U}$ is Donsker, for both the $Y^a$ and $Y^b$ distributions.

Third, this Donsker property requires two things: the complexity of $\mathcal{U}$ cannot be too great, and the envelope function must have certain finite moments.
Because utility functions are monotonic, Corollary 3.1 of \citet{vanderVaart1996} says their complexity is not too great, so the important consideration in practice is the finite moments.

In practice, this finite moment assumption means we should worry about using linear utility if the distribution is fat-tailed, but requiring even a small amount of concavity of the utility functions can fix the problem, as the following examples illustrate.
Alternatively, if there is a satiety point after which utility does not increase, then utility can be linear up to the point of satiety, even for fat-tailed distributions.

For example, consider CRRA utility $u_\theta(\cdot)$ with random variable $Y\in\mathcal{Y}=[1,\infty)$, and let $\mathcal{U}=\{u_\theta(\cdot) : c\le\theta\le d\}$.
The envelope function is $u_c(\cdot)$ because $u_c(y)\ge u(y)\ge0$ for any $y\in\mathcal{Y}$ and any $u\in\mathcal{U}$.
In this case, the ``finite moment'' assumption is $\E[u_c(Y)^2]<\infty$.
If $\E(Y^2)<\infty$, then $\E[u_c(Y)^2]\le\E[u_0(Y)^2]=\E[(Y-1)^2]<\infty$, so the assumption is satisfied for any $c\ge0$.
If instead $Y$ lacks a finite second moment, then linear utility ($\theta=0$) cannot be included, but relatively small $\theta$ still can be.
For example, let $Y=1+\abs{C}$, where $C$ follows a Cauchy distribution.
% https://www.wolframalpha.com/input/?i=integral+of+%282%2Fpi%29%2F%281%2B%28x-1%29%5E2%29+from+x%3D1+to+infinity
If $c=0$, then the envelope function is $u_0(y)=y-1$, so its second moment $\E[(Y-1)^2]=\E(C^2)$ is not finite, violating the assumption.
(Nor is the expected utility $\E(Y-1)$ finite, which suggests that people do not use linear utility and/or expected utility in this setting.)
% u <- function(x,th) (x^(1-th)-1)/(1-th)
% divergent = integrate(function(x) (u(x, 0 ))^2*(2/pi)/(1+(x-1)^2), 1, Inf)
% divergent = integrate(function(x) (u(x, 0.4 ))^2*(2/pi)/(1+(x-1)^2), 1, Inf)
% 11.3 = integrate(function(x) (u(x, 0.6 ))^2*(2/pi)/(1+(x-1)^2), 1, Inf)
% 1.61 = integrate(function(x) (log(x))^2*(2/pi)/(1+(x-1)^2), 1, Inf)
% 0.6 = integrate(function(x) (u(x, 1.5 ))^2*(2/pi)/(1+(x-1)^2), 1, Inf)
However, if instead $c=0.6$, then the envelope function's second moment is finite, equal to $11.3$.%
\footnote{R code: \code{integrate(function(x) ((x^(1-0.6)-1)/(1-0.6))^2*2*dt(x-1,1), 1, Inf)}
}
Alternatively, if instead of a $t$-distribution with one degree of freedom (Cauchy), $C$ follows a $t$-distribution with two degrees of freedom, then the second moment is finite even allowing $\theta$ very close to zero; e.g., the second moment is $100.7$ with $\theta=0.01$.%
\footnote{R code: \code{integrate(function(x) ((x^(1-0.01)-1)/(1-0.01))^2*2*dt(x-1,2), 1, Inf)}}

In the lower tail, there is a similar tradeoff between restrictions on $\theta$ and restrictions on the distribution of $Y$.
Continuing the CRRA example, there was no such problem above because $Y\ge1$ and $u_\theta(1)=0$ for any $\theta$; that is, the distribution restriction $Y\ge1$ was sufficient to avoid problems, regardless of $\theta$.
If instead $Y\in\mathcal{Y}=[0,\infty)$, then $u_\theta(y)\to-\infty$ as $y\downarrow0$ for $\theta\ge1$.
If we restrict $\theta\le d<1$, then there is no problem because $-\infty<u_d(0)\le u_\theta(0)\le-1$ for any $0\le\theta\le d$.
That is, the utility function restriction $d<1$ is sufficient to avoid lower tail problems, regardless of the distribution of $Y$.
In intermediate cases, there is a tradeoff, with larger $d$ allowed as the distribution of $Y$ places less probability near zero.

\section{Empirical Illustrations}
\label{sec:emp}

To illustrate the results and interpretations of the new methodology, I present various empirical examples using the shifted CRRA utility functions from \cref{sec:motivation}, which are also a special case of hyperbolic absolute risk aversion.
Replication code is in the supplementary material and (along with additional examples) on my website.%
\footnote{\Blinded{[BLINDED]}{\url{https://kaplandm.github.io}}}
All datasets are also publicly available, partly through the R \citep{R.core} package \texttt{wooldridge} \citep{R.wooldridge}, based on \citet{WooldridgeIntroText}.

Unless otherwise noted, $\alpha=0.05$ (one-sided) and $999$ bootstrap replications were used.

\subsection{Comparison with CDF-Based Inner CS}
\label{sec:emp-CDF}

For comparison, I show results from \citet{GoldmanKaplan2018c} that can be interpreted as an inner CS for the range of values on which the CDF of $Y^a$ lies below the CDF of $Y^b$.
That is, writing the two CDFs as $F_a(\cdot)$ and $F_b(\cdot)$, the population object of interest is the consensus set
\begin{equation*}
\mathcal{V} \equiv \{v : F_a(v)<F_b(v), v\in\R\} .
\end{equation*}
In the context of income or consumption, as in \citet{Atkinson1987}, this can be interpreted as 
the set of poverty lines $v$ for which distribution $Y^a$ has a lower poverty rate than $Y^b$.
A more general interpretation of $\mathcal{V}$ is the set of values $v$ for which $Y^a$ has a lower probability than $Y^b$ of a value below $v$ (or equivalently, higher probability of a value above $v$).
Parallel to $\hat{\mathcal{C}}_1$, the inner CS $\hat{\mathcal{V}}_1$ should satisfy $\Pr(\hat{\mathcal{V}}_1\subseteq\mathcal{V})\ge1-\alpha+o(1)$ for confidence level $1-\alpha$.

Method 5 of \citet[p.\ 153]{GoldmanKaplan2018c} provides an MTP with strong control of finite-sample FWER (Theorem 9, p.\ 155), which can be used to construct an inner CS parallel to \cref{eqn:CS-based-on-MTP,eqn:FWER-CP}.
Specifically, for $H_{0v}\colon F_a(v)\ge F_b(v)$, the inner CS is $\hat{\mathcal{V}}_1 = \{v : H_{0v}\textrm{ rejected}\}$, and the finite-sample coverage probability is
\begin{equation}
\Pr(\hat{\mathcal{V}}_1\subseteq\mathcal{V})
=\Pr(\textrm{no true }H_{0v}\textrm{ rejected})
=1-\overbrace{\Pr(\textrm{reject any true }H_{0v})}^{\equiv\FWER\le\alpha}
\ge 1-\alpha .
\end{equation}

\subsection{Investor Preferences over ETF Returns}
\label{sec:emp-ETF}

This example considers the set of investor preferences for which one exchange-traded fund (ETF) yields a better return distribution than another.
Specifically, I compare weekly returns of VTI (Vanguard's total US stock market index ETF) and BSV (Vanguard's short-term bond ETF).
Close prices for the 417 weeks of 2012--2019 are from Google Finance via Google Sheets.%
\footnote{\url{https://docs.google.com/spreadsheets/d/1ctjNbPyAaPfaLy3dp82TGFmpZ6afnQkAnWcOjJJKhk4}}
Given the nature of the two ETFs, qualitatively we expect more risk-averse investors (higher $\theta$ and $s$) to prefer BSV and expect less risk-averse investors (lower $\theta$ and $s$) to prefer VTI.
However, it is not obvious how ``more risk-averse'' and ``less risk-averse'' quantitatively map to the utility function space.

This example relates to the finance literature using stochastic dominance for portfolio evaluation and selection.
For example, the idea of ``stochastic dominance efficiency'' is that a particular portfolio is not dominated by any other portfolio within some class.
This can be used for portfolio evaluation \citep[e.g.,][]{Post2003,LintonEtAl2014} or portfolio selection \citep[e.g.,][]{DentchevaRuszczynski2006,HodderEtAl2015}.
However, dominance is a blunt concept.
For example, second-order stochastic dominance requires that a portfolio's return distribution is preferred given all possible risk-averse utility functions, even those that we do not think are realistic.
Thus, one portfolio may technically fail to dominate another, yet have higher expected utility for nearly all utility functions corresponding to real investors.
This is essentially the empirical finding below: although VTI does not stochastically dominate BSV, there is strong evidence (the inner confidence set) that the set of utility functions achieving higher expected utility with VTI includes almost all realistic utility functions, as informed by the literature estimating individuals' degree of risk aversion.

As described in \cref{sec:motivation}, the set of utility functions I consider is a special case of the hyperbolic absolute risk aversion (HARA) family characterized by Nobel Prize winner Robert C.~\citet{Merton1971}.
Specifically, his (43) is equivalent to mine (up to normalizing constants) after substituting notation $\theta=1-\gamma$ and $s=-(1-\gamma)\eta/\beta$ (and $y=C$), and restricting to non-increasing absolute risk aversion ($\theta\ge0$).
Quadratic utility is also commonly used in finance but primarily for convenience rather than economic reasons; even as early as 1974, the opening line of a paper \citep{Sarnat1974} says, ``The shortcomings of a quadratic utility function are so serious and so widely known that by now one might assume that it would simply have been dropped from consideration.''
Searching Google Scholar for ``HARA utility investment'' yields tens of thousands of results, including several thousand in the past few years.%
\footnote{\url{https://scholar.google.com/scholar?as_ylo=2019&q=hara+utility+investment}}

\begin{figure}[htbp]
\centering
\hfill
\includegraphics{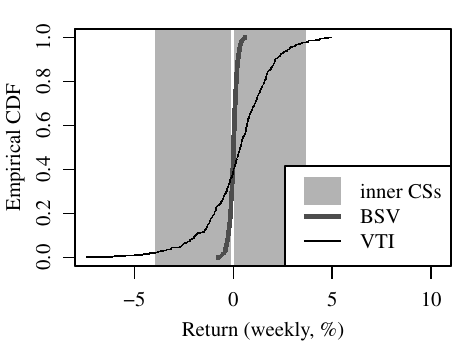}
\hfill
\includegraphics{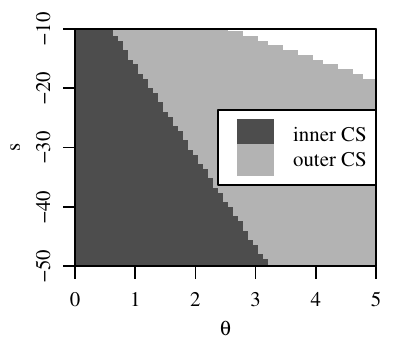}
\hfill\null
\caption{Weekly returns of BSV (short-term bond ETF) and VTI (US stock market ETF).}
\label{fig:ex-ETF}
\end{figure}

In a one-period buy-and-hold investment setting, applying the utility function to the ETF return itself is equivalent to applying it to wealth, other than a corresponding change in interpretation of the shift parameter $s$.
Letting random variable $R$ denote the ETF's return as a percentage, an investor's initial wealth $w$ becomes $w(1+R/100)$, and the investor prefers the ETF with higher $\E[u(w(1+R/100))]$.
Letting $\tilde{s}$ denote the subsistence level of wealth, and ignoring normalizing constants (denoted $\propto$), the utility function of wealth can be written in terms of return $R$:
\begin{align*}
u\bigl(w(1+R/100)\bigr)
  &\propto [w(1+R/100)-\tilde{s}]^{1-\theta}
% \\&
= \{ (w/100)[R - \overbrace{100(\tilde{s}/w - 1)}^{s}] \}^{1-\theta}
\\&= \overbrace{(w/100)^{1-\theta}}^{\textrm{constant}}
     \overbrace{(R-s)^{1-\theta}}^{\propto u(R)} ,
\end{align*}
where the shift parameter $s=100(\tilde{s}/w - 1)$ is the percentage change in initial wealth that would yield the subsistence level of wealth, like $s=-20$ if $\tilde{s}=0.8w$.
This is the interpretation of $s$ in \cref{fig:ex-ETF}.
That is, other than the interpretation of the shift parameters $s$, in this setting it is equivalent to maximize $\E[u(R)]$ as to maximize $\E[u(w(1+R/100))]$.
Below, I include some high wealth subsistence levels to show that only for such investors is there strong evidence of preference for the low-risk bond ETF over the stock ETF, and even then only if they have large risk aversion parameter $\theta$, too.

To inform the choice of possible $\theta$ values, I use the empirical literature estimating the coefficient of relative risk aversion.
\Citet[\S1]{Chetty2006} uses a life-cycle utility maximization model to characterize the coefficient of relative risk aversion in terms of labor supply elasticities estimated in the empirical literature, without assuming a particular utility functional form (and without assuming time separability), and including an upper bound for the coefficient regardless of consumption--labor complementarity.
He then uses elasticity estimates from a variety of economic fields and identification strategies to compute the corresponding coefficient of relative risk aversion in Table 1.
All values in columns (6)--(7) are in the interval $[0,2.5]$.
This includes studies of ``prime-age males, married women, retired individuals, and low-income families'' (p.~1829).
\Citet{Chetty2006} summarizes, ``The bottom line is that generating $\gamma$ significantly greater than $2$ would require complementarity and labor supply patterns that contradict evidence to date sharply'' (p.~1830).
In the working paper \citep{Chetty2003}, he notes that ``most economists believe based on introspection that $\gamma\in[1,5]$'' (p.~1) and discusses reasons why previous estimates are much noisier and likely biased; for example, see footnote 1 as well as the end of the top paragraph on page 3.
\Citet{ElminejadEtAl2022} reach a similar conclusion about other estimates in their meta-analysis of $92$ studies, including strong evidence of publication bias in favor of large estimates.
For example, their Figure 3 shows that all the highest-precision estimates (standard error below $0.1$) are in the range $[0,2]$, while most estimates above $5$ are very low precision.
While most individuals may have a coefficient of relative risk aversion in the interval $[0,2]$, there are likely some with higher values, and in this example it is interesting to see how much risk aversion is required to prefer the low-risk bond ETF, so I use the range $\theta\in[0,5]$.

\Cref{fig:ex-ETF}'s left panel shows the clear difference between the two empirical return distributions.
As expected, the VTI stock ETF's empirical distribution is much more dispersed than the BSV bond ETF's empirical distribution.
The shaded region below zero shows the $95\%$ inner CS for values at which the BSV CDF is below the VTI CDF, and the shaded region above zero shows the $95\%$ inner CS for values at which the VTI CDF is below the BSV CDF.
Despite the clear statistical meaning (that the CDFs cross), this provides relatively little economic insight.
Economically, the CDF analysis suggests that investors with a high enough degree of risk aversion do not prefer VTI over BSV, but it does not help us learn quantitatively about the set of investor preferences for which VTI is better.

\Cref{fig:ex-ETF}'s right panel shows the inner and outer $90\%$ confidence sets for the true set of utility functions receiving higher expected utility from the VTI stock ETF than from the BSV bond ETF.
Qualitatively, the results match our expectations: there is strong evidence that less risk-averse investors prefer VTI (inner CS) and that more risk-averse investors prefer BSV (outside the outer CS).
In the middle, there is not enough data to decide in favor of VTI or BSV at the $90\%$ confidence level.

Quantitatively, \cref{fig:ex-ETF} suggests almost all investors should prefer the stock ETF VTI over the bond ETF BSV.
Not only is the area outside the outer CS (strong evidence of preference for BSV over VTI) relatively small, but the corresponding $(\theta,s)$ values seem rare among investors in the real world.
First, recall the interpretation of $s=-20$ that the investor's subsistence level of wealth is $80\%$ of their initial wealth.
With $s\le-20$ (which likely everyone is), all $\theta\in[0,5]$ are in either the inner or outer CS in favor of VTI.
Further, with $s=-50$, all $\theta\in[0,3]$ are in the inner CS, and arguably almost everyone has even lower $s<-50$.
Combined with the aforementioned empirical evidence that $\theta\le2$, \cref{fig:ex-ETF} suggests that while some hypothetical investors (in the upper-right) prefer BSV over VTI, more realistic investors with $s<-35$ and $\theta\le2$ are all in the inner CS, meaning there is strong evidence that all such investors prefer VTI over BSV.

\subsection{Job Training}
\label{sec:emp-NSW}

I analyze data from two different job training programs.
In \cref{sec:emp-JTPA}, I use the same sample of Job Training Partnership Act (JTPA) data analyzed by \citet{AbadieEtAl2002} and \citet{KaplanSun2017}.
In this section, I use the National Supported Work (NSW) Demonstration data in the \texttt{wooldridge} package's \texttt{jtrain2} dataset; NSW offered ``training'' through subsidized real work experience.
In both cases, the training itself is the ``treatment.''
Individuals are randomly assigned to the treatment or control group, but those assigned to treatment may still decide not to get ``treated'' (trained).
That is, the assignment is randomized, but the actual treatment itself is not randomized due to non-compliance.
Consequently, instead of comparing treated and untreated individuals, I compare ``assigned'' (to treatment group) and ``not assigned'' (i.e., assigned to control group) individuals; that is, I examine the intention-to-treat effect on the earnings distribution.
This is different than comparing treated and untreated potential outcome distributions due to lack of perfect compliance, but intention-to-treat effects are still economically meaningful and commonly studied.

The CDF-based approach has a meaningful economic interpretation here.
The CDF-based consensus set $\mathcal{V}$ is the set of earnings values $v$ such that the treatment-assigned distribution has a higher probability of at least $v$ earnings than the not-assigned distribution.
Equivalently, the treatment-assigned distribution has a lower ``poverty'' rate for all $v\in\mathcal{V}$, defining ``poverty'' as having earnings below $v$.

The utility-based approach can be interpreted in terms of social welfare.
Tracing back to \citet[p.~257]{Atkinson1970}, the ``Atkinson social welfare function'' has the same form $\E[u(Y)]$ as expected utility but is interpreted in terms of the welfare of a society with distribution $Y$.
The parameter $\theta$ interpreted as risk aversion for an individual now represents inequality aversion.
For example, if $Y^a$ has both higher mean and higher variance (inequality) than $Y^b$, a more inequality-averse function may judge $Y^b$ to have higher social welfare in the sense of $\E[u(Y^b)]>\E[u(Y^a)]$, whereas a less inequality-averse function may instead judge $Y^a$ to have higher social welfare.
In the context of job training, the consensus set $\mathcal{C}$ is the set of functions $u(\cdot)$ for which the treatment-assigned earnings distribution $Y^a$ yields higher social welfare in the sense of $\E[u(Y^a)]>\E[u(Y^b)]$.

This particular social welfare function can also be viewed through the lens of \citet{Vickrey1945}, \citet{Harsanyi1953}, and \citet{Rawls1971}.
It answers the question: given expected utility maximization, would an individual with utility function $u(\cdot)$ prefer to be reincarnated as a random individual in a society with the treatment-assigned earnings distribution $Y^a$ or the not-assigned distribution $Y^b$?%
\footnote{\label{ft:OP}The ``veil of ignorance'' (or ``original position'') was popularized by Rawls, but the concept is found in earlier economics articles, combined with expected utility maximization: \citet{Vickrey1945} writes (\emph{emphasis} added), ``to \emph{maximize the aggregate of such utility over the population} is equivalent to choosing that distribution of income which such an individual would select were he asked which of various variants of the economy he would like to become a member of, assuming that once he selects a given economy with a given distribution of income he has an \emph{equal chance of landing in the shoes of each member of it}'' (p.\ 329); and \citet{Harsanyi1953} writes, ``Now, a value judgment on the distribution of income would show the required impersonality to the highest degree if the person who made this judgment\ldots had exactly the same chance of obtaining the first position (corresponding to the highest income) or the second or the third, etc., \ldots.  [Such judgments] may still be interpreted as an expression of what sort of society one would prefer if one had an equal chance of being `put in the place of' any particular member of the society, so that the cardinal utility `maximized' in value judgments concerning social welfare and the cardinal utility maximized in choices involving risk may be regarded as being fundamentally based upon the same principle'' (p.~435).}
This is not asking whether a specific existing individual should take the training, which depends only on that individual's utility function.
Instead, it asks a social question about which society as a whole is better: a society with this job training program available to everyone, or a society without it?
Then, we want to learn how broad a consensus (in terms of different ``utility'' functions $u(\cdot)$, or equivalently social welfare functions) agrees that the treatment-assigned distribution is superior, $\E[u(Y^a)]>\E[u(Y^b)]$.

Given the above interpretation and the empirical evidence noted in \cref{sec:emp-ETF}, I consider $\theta\in[0,3]$.
The range of shift values $s$ includes negative numbers because even with zero earnings an individual still has positive consumption.
More generally, as noted by \citet[p.~389]{Merton1971}, $s$ (proportional to his $-\eta$) alters how the degree of relative risk aversion changes with the level of earnings.

Despite the meaningful economic interpretations, there are important limitations of this particular empirical example, including the following.
First, the training program costs are not accounted for, which gives an unfair advantage to the training-assigned population.
Second, if the job training primarily acts as a signal to employers without increasing human capital much, then the effect of a large-scale implementation may be much smaller than the effect in the experiment, and similarly the long-run effect may be much smaller than the short-run effect.
Third, any ``effect'' is with respect to other employment services already available, which have certainly changed in the decades since.
Fourth, the ``treatment'' may not have been implemented the same across the many program sites (cities), but they are pooled together in my analysis.
Fifth, this comparison only considers two possible training-offer rules: offer nobody training, or offer everyone training.
In future work, it could be helpful to combine this paper's consideration of a set of utility (social welfare) functions with consideration of a set of possible treatment assignment rules, as in the empirical welfare maximization literature.

\begin{figure}[htbp]
\centering
\hfill
\includegraphics{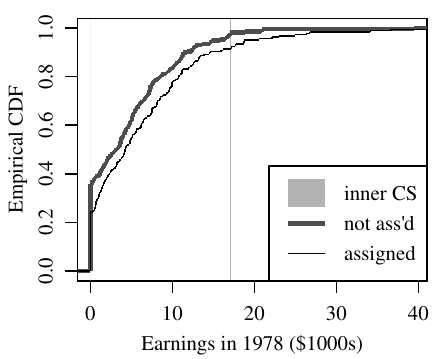}
\hfill
\includegraphics{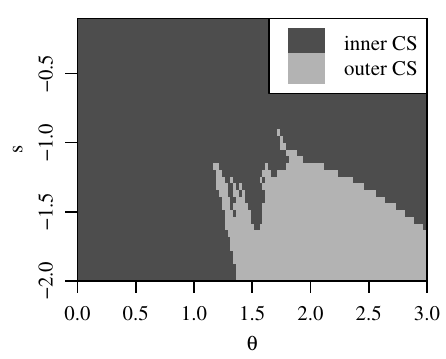}
\hfill\null
\caption{Earnings by job training.}
\label{fig:ex-jtrain2-train}
\end{figure}

\Cref{fig:ex-jtrain2-train} shows the NSW results, with $\alpha=0.025$.
The outcome variable is earnings in 1978 (in \USD{1000}s).
The left graph shows the training-assigned empirical CDF is below the not-assigned empirical CDF across a wide range of earnings values.
However, the CDF-based inner CS is very small because the vertical distance between empirical CDFs is not large at any given point.
(Even with $\alpha=0.1$, the inner CS remains small.)

\Cref{fig:ex-jtrain2-train}'s right graph shows the utility function (social welfare function) inner CS for the NSW data.
Contrasting the nearly-empty CDF-based inner CS, this inner CS includes most of the functions considered.
The inclusion of small $\theta$ (less inequality-averse) reflects the larger gains from training in the upper part of the distribution.
The inclusion of larger $\theta$ (more inequality-averse) reflects the large difference in the lower tail.
Although the inner CS does not extend to fill the lower-right area, reflecting greater statistical uncertainty, increasing from $\alpha=0.025$ to $\alpha=0.05$ enlarges the inner CS to cover all points in the graph.
Altogether, there is relatively strong evidence of consensus that the training-assigned earnings distribution is better, with the consensus spanning (at least) the shown range of social welfare function parameters $(\theta,s)$ used to define ``better.''

\section{Theory}
\label{sec:theory}

This section formally develops the theoretical results justifying the newly proposed methods.
\Cref{sec:limits} establishes the Gaussian limit of the expected utility difference process.
\Cref{sec:theory-MTP,sec:theory-CS} formally state MTP and CS methods and their properties.
\Cref{sec:band} introduces a uniform (over utility functions) confidence band for the expected utility difference.

The appendix has additional theoretical results used to justify these methods: \cref{sec:Donsker} has results characterizing Donsker classes of utility functions, and \cref{sec:theory-BS} establishes exchangeable bootstrap consistency.

The appendix also has additional methods and their properties: \cref{sec:stepdown} describes a stepdown procedure that improves the power of the MTP, and \cref{sec:test} describes hypothesis tests related to restricted stochastic dominance.

Shorter proofs are included below; longer proofs are in \cref{sec:app-pfs}.

For notation, to more closely match familiar conventions, I now write $f\in\mathcal{F}$ instead of $u\in\mathcal{U}$ and use the following.
Let $\{\Gn f: f\in\mathcal{F}\}$ denote the empirical process indexed by the set of functions $\mathcal{F}$, where $\Gn \equiv \sqrt{n}(\Pn-P)$ is the empirical process, $P$ is the true population measure, $\Pn=n^{-1}\sum_{i=1}^{n}\delta_{X_i}$ is the empirical measure (where $\delta$ is the Dirac delta function), and $Qf \equiv \int f\diff{Q}$
% =\E_{Y\sim Q}[f(Y)]$ 
for any measure $Q$, and let $\weaklyto$ denote weak convergence.
The two samples/populations are distinguished by superscript or subscript $a$ or $b$: $P^af=\E[f(Y^a)]$ for $Y^a\sim P^a$, $P^bf=\E[f(Y^b)]$ for $Y^b\sim P^b$, $\Pna f=n_a^{-1}\sum_{i=1}^{n_a}f(Y^a_i)$, $\Pnb f=n_b^{-1}\sum_{i=1}^{n_b}f(Y^b_i)$, and
\begin{equation}
\label{eqn:Gn}
\Gna \equiv \sqrt{n_a}(\Pna-P^a) ,\quad
\Gnb \equiv \sqrt{n_b}(\Pnb-P^b) ,\quad
\Gnd \equiv \sqrt{n_a}[(\Pna-\Pnb)-(P^a-P^b)]
.
\end{equation}
The statement $n\to\infty$ abbreviates $n_a,n_b\to\infty$.
Let $\ell^\infty(\mathcal{F})$ denote the space of all bounded functions $\mathcal{F}\mapsto\R$, equipped with norm $\norm{g}_{\mathcal{F}}=\sup_{f\in\mathcal{F}}\abs{g(f)}$.
Set difference $\setminus$ means $\mathcal{A}\setminus\mathcal{B}\equiv\{a:a\in\mathcal{A},a\notin\mathcal{B}\}$.

\subsection{Asymptotic Distributions}
\label{sec:limits}

\Cref{sec:limits-EUdiff} provides the Gaussian limit of the expected utility difference process (indexed by utility functions).
\Cref{sec:limits-suprema} extends this result to suprema.

\subsubsection{Limiting {Gaussian} Process for Expected Utility Difference}
\label{sec:limits-EUdiff}

\Cref{res:limit} states the Gaussian limit of the expected utility empirical process (indexed by utility functions), given the utility function class is Donsker and given \Cref{a:iid}.
\Cref{sec:Donsker} contains details (beyond \cref{sec:CS-justification}) about Donsker classes of utility functions, with formal results in \cref{res:Donsker-class-monotone,res:Donsker-class-CRRA}.
\Cref{a:iid} states there are two iid samples, independent of each other, whose sample sizes are of similar magnitude; it is similar to Assumption 2 (page 75) in \citet{BarrettDonald2003}, among others.

\begin{assumption}
\label{a:iid}
Random variables $Y^a_1,Y^a_2,\ldots$ are iid realizations of random variable $Y^a$ with distribution $P^a$.
Independent of the $Y^a_i$ are $Y^b_1,Y^b_2,\ldots$, iid realizations of $Y^b$ with distribution $P^b$.
The respective data samples are $Y^a_i$ for $i=1,\ldots,n_a$ and $Y^b_j$ for $j=1,\ldots,n_b$, where $n_a/n_b=\lambda+o(1)$ as $n_a,n_b\to\infty$.
\end{assumption}

\begin{theorem}
\label{res:limit}
Let $\mathcal{F}$ be a $P^a$-Donsker and $P^b$-Donsker class of utility functions.
Let \Cref{a:iid} hold, as well as the notation in \cref{eqn:Gn}.
Then,
$\Gna \weaklyto \Ga$ and $\Gnb \weaklyto \Gb$ in $\ell^\infty(\mathcal{F})$, 
where $\Ga$ and $\Gb$ are respectively $P^a$- and $P^b$-Brownian bridges, i.e., mean-zero Gaussian processes with covariance functions $\E[\Ga f \Ga g] =P^a fg - P^a f P^a g$ and $\E[\Gb f \Gb g] =P^b fg - P^b f P^b g$, respectively, and they are independent ($\Ga \independent \Gb$).
Further, the convergence is joint: $(\Gna,\Gnb)\weaklyto(\Ga,\Gb)$.
\end{theorem}
\begin{proof}
Because $\mathcal{F}$ is $P^a$-Donsker and $P^b$-Donsker, the marginal results are immediate; e.g., see (2.1.1) and (2.1.2) in \citet{vanderVaartWellner1996}.
There are only two processes, so the joint convergence follows from Theorem 1.4.8 of \citet{vanderVaartWellner1996}.
\end{proof}

\Cref{res:limit-diff} extends \cref{res:limit} to expected utility differences.

\begin{corollary}
\label{res:limit-diff}
Under the conditions of \cref{res:limit}, $\Gnd\weaklyto\Gd$ in $\ell^\infty(\mathcal{F})$,
where $\Gd$ is a mean-zero Gaussian process with covariance function
\begin{equation*}
 \E[\Gd f \Gd g]
=P^a fg - P^a f P^a g
 +\lambda^2
  [P^b fg - P^b f P^b g] .
\end{equation*}
\end{corollary}
\begin{proof}
Rearranging with algebra and applying \cref{res:limit},
\begin{equation*}
\overbrace{
\sqrt{n_a}
[(\Pna-\Pnb)-(P^a-P^b)]
}^{\Gnd}
= 
 \overbrace{
 \sqrt{n_a}(\Pna-P^a)
 }^{=\Gna\weaklyto\Ga}
-\overbrace{
 \sqrt{n_a/n_b}
 }^{\to\lambda}
 \overbrace{
 \sqrt{n_b}(\Pnb-P^b)
 }^{=\Gnb\weaklyto\Gb}
\weaklyto
\Gd
\equiv \Ga - \lambda\Gb ,
\end{equation*}
where the $\Delta$ superscript stands for ``difference.''
Both $\Ga$ and $\Gb$ are mean-zero, so $\Gd$ is, too.
Since additionally $\Ga \independent \Gb$, the covariance function is
\begin{align*}
\E[\Gd f \Gd g]
  &= \E[(\Ga - \lambda\Gb) f (\Ga - \lambda\Gb) g]
\\&= \E[\Ga f \Ga g]
 +\lambda^2
  \E[\Gb f \Gb g]
 -\lambda[\E(\Ga f \Gb g)+\E(\Gb f \Ga g)]
\\&= P^a fg - P^a f P^a g
 +\lambda^2
  [P^b fg - P^b f P^b g]
\\&\quad
 -\lambda[\overbrace{\E(\Ga f)}^{=0}\overbrace{\E(\Gb g)}^{=0} + \overbrace{\E(\Gb f)}^{=0} \overbrace{\E(\Ga g)}^{=0}] .
\qedhere
\end{align*}
\end{proof}

\subsubsection{Distributions of Suprema}
\label{sec:limits-suprema}

The distributions of maximal $t$-statistics are useful for inference.
Denote the pointwise (scalar) variance of $\Gd f$ as
\begin{equation}
\label{eqn:sigma-f}
\sigma^2_f
\equiv \Var(\Gd f)
= P^a f^2 - (P^af)^2
 +\lambda^2
  [P^b f^2 - (P^bf)^2] ,
\end{equation}
with $\hat\sigma_f$ an estimator like \cref{eqn:sigma-hat-f-BS}.
For $\mathcal{S}\subseteq\mathcal{F}$, using notation from \cref{eqn:Gn,res:limit-diff}, define
\begin{align}
\label{eqn:T}
T_f &\equiv \Gd f / \sigma_f , \quad
T^{\mathcal{S}\vee}
\equiv \sup_{f\in\mathcal{S}}
T_f ,\quad
T^{\mathcal{S}\wedge}
\equiv \inf_{f\in\mathcal{S}}
T_f ,\quad
\abs{T}^{\mathcal{S}\vee} \equiv \sup_{f\in\mathcal{S}}\abs{T_f}
, \\
\label{eqn:T-hat}
\hat{T}_f &\equiv \Gnd f / \hat\sigma_f , \quad
\hat{T}^{\mathcal{S}\vee} \equiv \sup_{f\in\mathcal{S}}\hat{T}_f , \quad
\hat{T}^{\mathcal{S}\wedge} \equiv \inf_{f\in\mathcal{S}}\hat{T}_f , \quad
\abs{\hat{T}}^{\mathcal{S}\vee} \equiv \sup_{f\in\mathcal{S}} \abs{\hat{T}_f} .
\end{align}
The corresponding limit distributions are given in \cref{res:sup}, under \Cref{a:sigma-f,a:sigma-hat-f}.
\Cref{a:sigma-hat-f} is satisfied by the bootstrap estimator in \cref{eqn:sigma-hat-f-BS}, among other possibilities.

\begin{assumption}
\label{a:sigma-f}
The pointwise variances $\sigma^2_f$ from \cref{eqn:sigma-f} are uniformly (over $f\in\mathcal{F}$) bounded away from zero.
\end{assumption}

\begin{assumption}
\label{a:sigma-hat-f}
The estimators $\hat\sigma_f$ are uniformly consistent: $\sup_{f\in\mathcal{F}}\abs{\hat\sigma_f-\sigma_f}\pconv0$.
\end{assumption}

\begin{corollary}
\label{res:sup}
Let the conditions for \cref{res:limit-diff} hold along with \Cref{a:sigma-f,a:sigma-hat-f}.
Then,
\begin{equation*}
\hat{T}_f \dconv T_f \sim \NormDist(0,1) 
\textrm{ for each }f\in\mathcal{F}
,\quad
\hat{T}^{\mathcal{S}\vee} \dconv T^{\mathcal{S}\vee} ,\quad
\hat{T}^{\mathcal{S}\wedge} \dconv T^{\mathcal{S}\wedge} ,\quad
\abs{\hat{T}}^{\mathcal{S}\vee} \dconv \abs{T}^{\mathcal{S}\vee} .
\end{equation*}
\end{corollary}
\begin{proof}
This follows from $\hat{T}_f\equiv \Gnd f/\hat\sigma_f=\Gnd f/\sigma_f+o_p(1)$ uniformly over $f\in\mathcal{F}$ (by \cref{a:sigma-f,a:sigma-hat-f}) and the continuous mapping theorem applied to \cref{res:limit-diff}.
\end{proof}

\subsection{Multiple Testing}
\label{sec:theory-MTP}

Formal statements of the multiple testing procedure and its properties are given below.

Writing the latter part of \cref{eqn:H0} in the new notation, let
\begin{equation}
\label{eqn:H0f}
H_{0f} \colon P^af-P^bf \le 0 , \quad
H_{1f} \colon P^af-P^bf > 0 , \quad
\textrm{for each } f\in\mathcal{F} .
\end{equation}
That is, $H_{0f}$ is true if $Y^a$ has lower expected utility (given $f$) than $Y^b$, whereas $H_{1f}$ is true (and $H_{0f}$ false) if instead $Y^a$ has higher expected utility.
Define the set of utility functions for which $Y^b$ is preferred (so the null is true) as
\begin{equation}
\label{eqn:F-T}
\mathcal{F}_T \equiv \{ f : P^af-P^bf\le0 \}
= \{ f : H_{0f}\textrm{ is true} \}
.
\end{equation}
As in \citet[\S9.1]{LehmannRomano2005text}, the familywise error rate (FWER) is
\begin{equation}
\label{eqn:FWER}
\FWER \equiv \Pr( \textrm{reject any }H_{0f}\textrm{ with }f\in\mathcal{F}_T ) ,
\end{equation}
and strong control of FWER at level $\alpha$ means $\FWER\le\alpha$ given any $\mathcal{F}_T$ (as opposed to ``weak control'' for only $\mathcal{F}_T=\mathcal{F}$).

\Cref{res:MTP-basic} shows the FWER control of the MTP in \cref{meth:MTP-basic}.
As usual, there are many possible ways to allocate the pointwise size while controlling FWER; \cref{meth:MTP-basic} has the same pointwise asymptotic size for each $H_{0f}$, which provides a reasonable default.

\begin{method}
\label{meth:MTP-basic}
Reject $H_{0f}$ when $\hat{T}_f>\tilde{T}^{\mathcal{F}\vee}_{1-\alpha}$, where $\hat{T}_f$ is from \cref{eqn:T-hat}, and $\tilde{T}^{\mathcal{F}\vee}_{1-\alpha}$ is the bootstrap critical value computed by \cref{meth:BScv} (with $\mathcal{S}=\mathcal{F}$).
\end{method}

\begin{proposition}
\label{res:MTP-basic}
Given \cref{res:sup,res:BS}, \cref{meth:MTP-basic} has strong control of asymptotic FWER at level $\alpha$.
\end{proposition}

\subsection{Confidence Sets}
\label{sec:theory-CS}

As in \cref{eqn:emp-consensus-set}, the consensus set $\mathcal{C}$ contains all the utility functions for which $Y^a$ is preferred.
In the new notation,
\begin{equation}
\label{eqn:C}
\mathcal{C}
\equiv
\{ f\in\mathcal{F}
  : P^a f - P^b f > 0 \} .
\end{equation}

\Cref{res:CS} states the asymptotic coverage properties of the confidence sets described in \cref{meth:CS}.
The inner and outer CSs are the same as before.
Additionally, a joint CS pair is described in terms of a uniform confidence band (\cref{sec:band}).
The joint CS pair satisfies
\begin{equation}
\label{eqn:CS-asy-joint-CP}
\limn \Pr(\hat{\mathcal{C}}_1\subseteq\mathcal{C}\subseteq\hat{\mathcal{C}}_2)
\ge 1-\alpha .
\end{equation}

\begin{method}
\label{meth:CS}
Inner CS: run an MTP of $H_{0f}\colon P^af-P^bf\le0$ over $f\in\mathcal{F}$ with strong control of FWER at level $\alpha$ (like \cref{meth:MTP-basic} or \cref{meth:stepdown}); 
then, $\hat{\mathcal{C}}_1=\{f\in\mathcal{F}:\textrm{MTP rejects }H_{0f}\}$.
Outer CS: run an MTP of $H_{0f}\colon P^af-P^bf\ge0$ and let $\hat{\mathcal{C}}_2=\{f\in\mathcal{F}:\textrm{MTP does not reject }H_{0f}\}$.
Joint/two-sided CS pair: 
compute two-sided uniform confidence band $\hat{b}_1(\cdot)$ and $\hat{b}_2(\cdot)$ using \cref{meth:band}, then let $\hat{\mathcal{C}}_1=\{f\in\mathcal{F}:\hat{b}_1(f)>0\}$ and $\hat{\mathcal{C}}_2=\{f\in\mathcal{F}:\hat{b}_2(f)>0\}$.
\end{method}

\begin{proposition}
\label{res:CS}
Given \cref{eqn:C} and \cref{res:MTP-basic,res:stepdown}, the inner and outer CSs in \cref{meth:CS} satisfy \cref{eqn:inner-CS-asyCP,eqn:outer-CS-asyCP}.
Given \cref{eqn:C} and \cref{res:band}, the joint CS pair in \cref{meth:CS} satisfies \cref{eqn:CS-asy-joint-CP}.
\end{proposition}

\subsection{Uniform Confidence Bands}
\label{sec:band}

This section considers uniform (over $f\in\mathcal{F}$) confidence bands for the expected utility difference $P^af-P^bf$.
Let $\hat{b}_1\colon\mathcal{F}\mapsto\R$ and $\hat{b}_2\colon\mathcal{F}\mapsto\R$ denote the lower and upper confidence band functions.
The goal is uniform $1-\alpha$ asymptotic coverage:
\begin{equation}
\label{eqn:band-CP}
1-\alpha =
\limn \Pr\{ \hat{b}_1(f) \le P^af-P^bf \le \hat{b}_2(f) \textrm{ for all }f\in\mathcal{F} \} .
\end{equation}

Such bands contain the most information among all inference methods proposed in this paper, but they are also the most difficult to comprehend (and thus communicate).
\Cref{meth:MTP-basic,meth:CS} can both be characterized as summaries of the uniform confidence band. %...and meth:test-SD-t
Such summaries improve comprehension at the cost of information loss.
At one extreme of the information--comprehension spectrum, the uniform band consists of a range of values for each $f\in\mathcal{F}$; at the other extreme, the tests in \cref{sec:test} consist of only a binary decision (reject or not).
The multiple testing procedure and confidence sets provide a balance, improving comprehension while retaining information across $f\in\mathcal{F}$.

\Cref{meth:band} constructs uniform bands whose asymptotic coverage is in \cref{res:band}.

\begin{method}[uniform confidence bands]
\label{meth:band}
First, run \cref{meth:BScv} to compute the bootstrap critical values $\abs{\tilde{T}}^{\mathcal{F}\vee}_{1-\alpha}$, $\tilde{T}^{\mathcal{F}\vee}_{1-\alpha}$, $\tilde{T}^{\mathcal{F}\wedge}_{\alpha}$, $\tilde{T}^{\mathcal{F}\vee}_{1-\alpha/2}$, and/or $\tilde{T}^{\mathcal{F}\wedge}_{\alpha/2}$, with the desired $\alpha$ and $\mathcal{S}=\mathcal{F}$.
Define $\hat\sigma_f$ as in \cref{eqn:sigma-hat-f-BS}.
For a symmetric two-sided uniform $1-\alpha$ confidence band,
\begin{equation}
\label{eqn:band-symm}
\hat{b}_1(f) =
(\Pna-\Pnb)f 
- \abs{\tilde{T}}^{\mathcal{F}\vee}_{1-\alpha}
  \hat\sigma_f / \sqrt{n_a} ,
\quad
\hat{b}_2(f) =
(\Pna-\Pnb)f 
+ \abs{\tilde{T}}^{\mathcal{F}\vee}_{1-\alpha}
  \hat\sigma_f / \sqrt{n_a} .
\end{equation}
For a one-sided uniform $1-\alpha$ confidence band, either
\begin{equation}
\label{eqn:band-1s}
\begin{split}
\hat{b}_1(f) &=
(\Pna-\Pnb)f 
- \tilde{T}^{\mathcal{F}\vee}_{1-\alpha}
  \hat\sigma_f / \sqrt{n_a} ,
\quad
\hat{b}_2(f) = \infty ,
\quad\textrm{or}\\
\hat{b}_2(f) &=
(\Pna-\Pnb)f 
- \tilde{T}^{\mathcal{F}\wedge}_{\alpha}
  \hat\sigma_f / \sqrt{n_a}
,\quad
\hat{b}_1(f) = -\infty .
\end{split}
\end{equation}
For an ``equal-tailed'' two-sided band (that may be conservative),
\begin{equation}
\label{eqn:band-2tail}
\hat{b}_1(f) =
(\Pna-\Pnb)f 
- \tilde{T}^{\mathcal{F}\vee}_{1-\alpha/2}
  \hat\sigma_f / \sqrt{n_a} ,
\quad
\hat{b}_2(f) =
(\Pna-\Pnb)f 
- \tilde{T}^{\mathcal{F}\wedge}_{\alpha/2}
  \hat\sigma_f / \sqrt{n_a} .
\end{equation}
\end{method}

\begin{proposition}
\label{res:band}
Given \cref{res:BS}, the bands in \cref{eqn:band-symm,eqn:band-1s} in \cref{meth:band} have exact asymptotic uniform coverage probability (i.e., satisfy \cref{eqn:band-CP}), and the band in \cref{eqn:band-2tail} has asymptotic uniform coverage probability of at least $1-\alpha$.
\end{proposition}

\section{Simulation}
\label{sec:sim}

Complementing the asymptotic theory in \cref{sec:theory}, the following simulation reports finite-sample coverage probabilities.
The inner and outer confidence sets are examined, along with a uniform confidence band.
Replication code in R \citep{R.core} is in the supplementary material and on my website.%
\footnote{\Blinded{[BLINDED]}{\url{https://kaplandm.github.io}}}
Along with other details, the numbers of simulation and bootstrap replications are in the table notes.

In the data-generating process, both $Y^a$ and $Y^b$ are log-normal with respective parameters $(\mu_a,\sigma_a)$ and $(\mu_b,\sigma_b)$.
The value of $(\mu_b,\sigma_b)$ varies as shown in the table, whereas $(\mu_a,\sigma_a)=(0,1)$ in every case.
Sampling is iid, with sample sizes $n_a$ (for $Y^a$) and $n_b$ (for $Y^b$) shown in the table.

I consider utility functions
\begin{equation}\label{eqn:CRRA-shifted}
\mathcal{F}=\{f_\theta:0\le\theta\le3\} , \quad
f_\theta(y) = \left\{
\begin{matrix}
\ln(y+0.1) & \textrm{if }\theta=1, \\
\frac{(y+0.1)^{1-\theta}-1}{1-\theta} & \textrm{if }\theta\ne1.
\end{matrix}
\right.
\end{equation}
This is equivalent to shifting $Y^a$ and $Y^b$ up by $0.1$ and using the usual constant relative risk aversion (CRRA) utility function.
Without the small shift, $\lim_{y\downarrow0}f_\theta(y)=-\infty$ for $\theta\ge1$, which naturally causes problems in finite samples.
The small shift makes the envelope function of $\mathcal{F}$ bounded from below (though not above).
For computation, the grid $\theta=0.0,0.1,\ldots,3.0$ is used.

The following coverage probability (CP) results are reported.
The nominal confidence level is $1-\alpha=0.9$.
The true set $\mathcal{C}$ is equivalently expressed in terms of the risk aversion parameter, as $\{\theta:f_\theta\in\mathcal{C}\}$.
If this is the empty set $\{ \}$, then the inner confidence set can still ``cover'' by also being the empty set.
For each simulated dataset, the exchangeable bootstrap uniform confidence band for $\Delta(f) \equiv P^af-P^bf$ (over $f\in\mathcal{F}$) is computed, using \cref{meth:band}.
The reported CP is the proportion of simulation replications (datasets) in which the band contained $P^af-P^bf$ for all $f\in\mathcal{F}$, i.e., in which $\hat{b}_1(f) \le \Delta(f) \le \hat{b}_2(f)$ for all $f\in\mathcal{F}$.
From this band, the (joint) inner confidence set $\hat{\mathcal{C}}_1$ and outer confidence set $\hat{\mathcal{C}}_2$ are computed, using \cref{meth:CS}.
The column ``both sets'' shows the proportion of replications in which $\hat{\mathcal{C}}_1 \subseteq \mathcal{C} \subseteq \hat{\mathcal{C}}_2$, which should be at least as big as the nominal $1-\alpha$ (recall the $\ge$ in \cref{res:CS}).
For additional detail, this is further split into simulated CP of $\hat{\mathcal{C}}_1 \subseteq \mathcal{C}$ and of $\mathcal{C} \subseteq \hat{\mathcal{C}}_2$; note that the sum of these two simulated probabilities always equals one plus the ``both sets'' simulated probability.

\begin{table}[htb]
\centering
\caption{\label{tab:sim-CP}Simulated coverage probability.}
\sisetup{round-precision=3}
\begin{threeparttable}
\begin{tabular}[c]{rr 
S[table-format=1.1,round-precision=1]S[table-format=-1.1,round-precision=1]
c
S[table-format=1.3]S[table-format=1.3]S[table-format=1.3]S[table-format=1.3]}
\toprule
 & & & & & \multicolumn{4}{c}{Probability of} \\
\cmidrule{6-9}
\multicolumn{1}{c}{$n_a$} & \multicolumn{1}{c}{$n_b$} & 
\multicolumn{1}{c}{$\sigma_b$}&\multicolumn{1}{c}{$\mu_b$}&
\multicolumn{1}{c}{$\{\theta:f_\theta\in\mathcal{C}\}$}&
\multicolumn{1}{c}{$\hat{b}_1 \le \Delta \le \hat{b}_2$} & \multicolumn{1}{c}{$\hat{\mathcal{C}}_1\subseteq\mathcal{C}\subseteq\hat{\mathcal{C}}_2$} & \multicolumn{1}{c}{$\hat{\mathcal{C}}_1\subseteq\mathcal{C}$} & \multicolumn{1}{c}{$\mathcal{C}\subseteq\hat{\mathcal{C}}_2$} \\
\midrule
  40 &   40 & 0.7 & -0.3 & [0.0, 2.8] & 0.860 & 0.954 & 0.962 & 0.992 \\
  40 &   40 & 0.7 &  0.0 & [0.0, 1.1] & 0.852 & 0.966 & 0.985 & 0.981 \\
  40 &   40 & 0.7 &  0.3 &    $\{ \}$ & 0.839 & 0.998 & 0.998 & 1.000 \\[2pt]
  40 &   40 & 1.0 & -0.3 & [0.0, 3.0] & 0.915 & 0.999 & 1.000 & 0.999 \\
  40 &   40 & 1.0 &  0.0 &    $\{ \}$ & 0.927 & 0.963 & 0.963 & 1.000 \\
  40 &   40 & 1.0 &  0.3 &    $\{ \}$ & 0.922 & 0.994 & 0.994 & 1.000 \\[2pt]
  40 &   40 & 1.3 & -0.3 & [0.2, 3.0] & 0.896 & 0.965 & 0.968 & 0.997 \\
  40 &   40 & 1.3 &  0.0 & [1.2, 3.0] & 0.875 & 0.972 & 0.988 & 0.984 \\
  40 &   40 & 1.3 &  0.3 & [2.5, 3.0] & 0.859 & 0.957 & 0.993 & 0.964 \\[4pt]
 100 &  100 & 0.7 & -0.3 & [0.0, 2.8] & 0.893 & 0.961 & 0.971 & 0.990 \\
 100 &  100 & 0.7 &  0.0 & [0.0, 1.1] & 0.881 & 0.972 & 0.990 & 0.982 \\
 100 &  100 & 0.7 &  0.3 &    $\{ \}$ & 0.898 & 0.999 & 0.999 & 1.000 \\[2pt]
 100 &  100 & 1.0 & -0.3 & [0.0, 3.0] & 0.929 & 1.000 & 1.000 & 1.000 \\
 100 &  100 & 1.0 &  0.0 &    $\{ \}$ & 0.914 & 0.957 & 0.957 & 1.000 \\
 100 &  100 & 1.0 &  0.3 &    $\{ \}$ & 0.915 & 1.000 & 1.000 & 1.000 \\[2pt]
 100 &  100 & 1.3 & -0.3 & [0.2, 3.0] & 0.883 & 0.969 & 0.977 & 0.992 \\
 100 &  100 & 1.3 &  0.0 & [1.2, 3.0] & 0.890 & 0.977 & 0.982 & 0.995 \\
 100 &  100 & 1.3 &  0.3 & [2.5, 3.0] & 0.872 & 0.953 & 0.988 & 0.965 
\\
\bottomrule
\end{tabular}
\begin{tablenotes}[para,flushleft]
\footnotesize{}
  Nominal level $1-\alpha=0.9$ for band and ``two-sided'' confidence set; $\mu_a=0$, $\sigma_a=1$, $\numnornd{1000}$ simulation replications, $\numnornd{999}$ bootstrap draws, $\theta=0,0.1,\ldots,3$, $f_\theta(y)=[(y+0.1)^{1-\theta}-1]/(1-\theta)$. In the header, $\Delta=P^a-P^b$, and $\hat{b}_1\le\Delta\le\hat{b}_2$ means $\hat{b}_1(f)\le P^af-P^bf \le\hat{b}_2(f)$ for all $f\in\mathcal{F}$, with $\mathcal{F}$ from \cref{eqn:CRRA-shifted}.
% Time difference of 2.353292 hours
\end{tablenotes}
\end{threeparttable}
\end{table}

\Cref{tab:sim-CP} shows the confidence band performing reasonably across DGPs, even with modest sample sizes.
With $n_a=n_b=40$ observations per sample, CP ranges from $0.839$ to $0.927$, depending on the parameters.
With $n_a=n_b=100$, CP improves, with all values in the range $[0.872,0.929]$.

\Cref{tab:sim-CP} also shows that the CS coverage probabilities are conservative, as expected (per the $\ge$ in \cref{res:CS}).
They are always at least $0.95$ (even though $1-\alpha=0.9$), sometimes $1.000$ (up to simulation error).
The confidence sets are all valid, but there is room for improvement in future work, like constructing $\hat{\mathcal{C}}_1$ by inverting the stepdown MTP in \cref{meth:stepdown}.

\section{Conclusion}

I have considered learning about the consensus set of utility functions for which one distribution is preferred to another (higher expected utility), which opens other areas for future research.
For example, extending the theory to non-iid sampling would be valuable, especially for the complex sampling designs often used for income, wealth, and consumption surveys.
Also valuable would be extensions to increase the flexibility of a class of utility functions while retaining computational feasibility and economic interpretation, whether inspired by economic theory or convenient basis functions or both, or extensions to frameworks beyond expected utility, such as utility functions that incorporate social inequality in addition to the individual's well-being.
Power could probably be improved through pre-testing or using other strategies from the (moment) inequality testing literature.
Other extensions include simultaneous ranking of more than two distributions or more than one pair of distributions.
For the former, it may be helpful to combine my approach with that of \citet{MogstadEtAl2023}, who consider ranking many distributions based on a scalar summary statistic.
My general approach can also be considered for ranking distributions using inequality measures indexed by utility functions or other parameters, or for stochastic monotonicity.
It can also extend to utility functions of multiple variables (like income and health).
Finally, this approach could be considered within choice models or in light of observed choices.
For example, if it is known that an individual prefers the first distribution, then the set of utility functions for which the first distribution has higher expected utility can be interpreted as the identified set for that individual's utility function, and the outer confidence set is a confidence set for the identified set.

\section*{Supplementary Materials}

% The supplementary appendix has additional (more technical) theoretical results, additional methodology, proofs not found in the main text, and an example algorithm to compute bootstrap critical values.
% Also provided is R code to replicate the simulation and empirical results.
R code is provided to replicate the simulation and empirical results.

\section*{Acknowledgments}

Thanks to Tim Armstrong for the initial idea, and to the Cowles Foundation more generally for their hospitality.
Thanks to Alyssa Carlson for helpful comments on multiple drafts, and to Saku Aura and other colleagues for feedback at a brown bag seminar.
I am also grateful for the helpful feedback from seminars at The University of Chicago, Boston University, and UC Santa Cruz (especially from St{\'e}phane Bonhomme, Jim Heckman, Azeem Shaikh, Alex Torgovitsky, Hiro Kaido, Iv{\'a}n Fern{\'a}ndez-Val, JJ Forneron, Pierre Perron, Jessie Li, and Juli{\'a}n Mart{\'i}nez-Iriarte) and the 2021 North American Summer Meeting of the Econometric Society (especially Ruli Xiao, Jackson Bunting, Yuya Sasaki, and Takuya Ura), as well as feedback from anonymous reviewers, the associate editor, and editor Ivan Canay, all of which helped improve this work.

\section*{Conflict of Interest Statement}

I have no competing interests to declare.

\singlespacing
\bibliographystyle{jpe}
% \bibliography{_bib}

\paperspacing

\newpage
\appendix
\paperspacing
\singlespacing

% \Blinded{}{
% \setcounter{page}{1}
% {\centering\LARGE%\Large
% Appendix for\\
% ``Inference on Consensus Ranking of Distributions''\\[12pt]
% \large%\normalsize
% David M.\ Kaplan\\
% \today

% }
% } %end unblinded

\section{Additional Theoretical Results}
\label{sec:app-theory}

This section includes additional theoretical results.
Besides justifying the methods proposed in the main text, \cref{sec:Donsker} may also be of independent interest.

\subsection{Donsker Classes of Utility Functions}
\label{sec:Donsker}

This section considers the key technical condition for applying empirical process theory to expected utility: whether the specified set of utility functions is Donsker.
\Cref{sec:Donsker-monotone} shows this holds very generally.
\Cref{sec:Donsker-CRRA} establishes the Donsker property of the popular but simple class of constant relative risk aversion (CRRA) utility functions under a slightly weaker envelope function condition.
A practical discussion of the finite moment assumption is in \cref{sec:CS-justification}.

\subsubsection{A large nonparametric class of utility functions}
\label{sec:Donsker-monotone}

Corollary 3.1 of \citet{vanderVaart1996} establishes that the class of essentially all utility functions is Donsker, subject to a common lower (or upper) bound and an envelope function with $2+\delta$ moments.
Below, the lower bound is stated as zero without loss of generality, because adding a constant to a utility functions does not affect the expected utility difference when comparing two distributions.
That is, if utility function $f(x)=\tilde{f}(x)+c$ for constant $c$, and $Y^a$ and $Y^b$ are random variables, then
\[ \E[f(Y^a)]-\E[f(Y^b)]
  =\E[\tilde{f}(Y^a)+c]
   -\E[\tilde{f}(Y^b)+c]
  =\E[\tilde{f}(Y^a)]
   -\E[\tilde{f}(Y^b)]. \]

\begin{lemma}[\citet{vanderVaart1996}, Cor.\ 3.1]
\label{res:Donsker-class-monotone}
The set $\mathcal{F}$ of non-decreasing utility functions $f \colon \R \mapsto \R_{\ge0}$ is $P$-Donsker if $PF^{2+\delta}<\infty$ for some $\delta>0$, where $F$ is an envelope function satisfying 
$0\le f\le F$ for all $f\in\mathcal{F}$.
\end{lemma}
\begin{proof}
This is Corollary 3.1 of \citet{vanderVaart1996}.
\end{proof}

\subsubsection{A parametric utility function class}
\label{sec:Donsker-CRRA}

Consider the simple but popular CRRA utility
\begin{equation}
\label{eqn:CRRA}
f_\theta(x) =
\left\{
\begin{matrix*}[l]
\dfrac{x^{1-\theta}-1}{1-\theta} & \textrm{if }\theta\ge0\textrm{ and }\theta\ne1 , \\[10pt]
\ln(x) & \textrm{if }\theta=1 .
\end{matrix*}
\right.
\end{equation}
The parameter $\theta$ represents risk aversion.
With $\theta=0$, $f_0(x)=x-1$, which is risk-neutral.
With $\theta>0$, $f_\theta(x)$ is concave in $x$, which is risk-averse.

\Cref{res:Donsker-class-CRRA} states that the VC dimension is one.
For any $\theta$, $f_\theta(1)=0$.
For any other $x\ne1$, $f_\theta(x)$ is decreasing in $\theta$; that is, $f_b(\cdot)\le f_a(\cdot)$ for any $b\ge a$.
Thus, the subgraphs are nested, so the VC dimension is $1$.
The Donsker property follows if the envelope function is square integrable, or if the somewhat weaker condition in Theorem 2.6.8 of \citet{vanderVaartWellner1996} holds.

\begin{lemma}
\label{res:Donsker-class-CRRA}
Given \cref{eqn:CRRA}, let $\mathcal{F}=\{f_\theta : 0\le\theta<\bar\theta\}$, allowing $\bar\theta=\infty$.
The VC dimension of $\mathcal{F}$ is one, and it is $P$-Donsker if $PF^2<\infty$, where $F$ is the envelope function.
\end{lemma}

\subsection{Exchangeable Bootstrap Consistency}
\label{sec:theory-BS}

``The'' exchangeable bootstrap consistently estimates the limiting Gaussian process $\Gd$ from \cref{res:limit-diff}.
This provides an alternative to simulating from a version of $\Gd$ with explicitly estimated covariance function.
Special cases of exchangeable bootstrap include the empirical bootstrap (multinomial weights), Bayesian bootstrap, $m$-out-of-$n$ bootstrap, and subsampling.
See Theorem 3.6.13 and equation (3.6.8) of \citet[p.\ 354--355]{vanderVaartWellner1996} for conditions on measurability and general weights.
The conditions on the weights are stated in \Cref{a:weights}.

\begin{assumption}
\label{a:weights}
For $n=n_a$ or $n=n_b$, let $(\tilde{W}_{n1},\ldots,\tilde{W}_{nn})$ denote the exchangeable random vector of nonnegative weights, independent of the data.
Denote the average $\bar{W}_n\equiv n^{-1}\sum_{i=1}^{n}\tilde{W}_{ni}$.
Further, as in (3.6.8) of \citet[p.\ 354]{vanderVaartWellner1996},
% \norm{}_{2,1} defined on page 177
\begin{gather*}
\sup_n \norm{\tilde{W}_{n1}-\bar{W}_n}_{2,1} 
\equiv \sup_n \int_{0}^{\infty}\sqrt{\Pr(\abs{\tilde{W}_{n1}-\bar{W}_n}>x)}\diff{x} < \infty , \\
n^{-1/2} \E \max_{1\le i\le n} \abs{\tilde{W}_{ni}-\bar{W}_n} \pconv 0 ,\quad
n^{-1} \sum_{i=1}^{n}(\tilde{W}_{ni}-\bar{W}_n)^2 \pconv c^2 > 0 .
\end{gather*}
\end{assumption}

Notationally, denote the bootstrap empirical process (difference) adjusted for $c$ as
\begin{equation}
\label{eqn:GndBS}
\begin{split}
\GndBS &\equiv \sqrt{n_a}[(\PnaBS-\PnbBS)-(\bar{W}^a\Pna-\bar{W}^b\Pnb)] / c , \\
\PnaBS &\equiv \frac{1}{n_a}\sum_{i=1}^{n_a}\tilde{W}^a_i \delta_{Y^a_i} , \quad
\PnbBS \equiv \frac{1}{n_b}\sum_{i=1}^{n_b}\tilde{W}^b_i \delta_{Y^b_i} ,
\end{split}
\end{equation}
where $(\tilde{W}^a_1,\ldots,\tilde{W}^a_{n_a})$ and $(\tilde{W}^b_1,\ldots,\tilde{W}^b_{n_b})$ are the exchangeable random vectors of bootstrap weights with respective averages $\bar{W}^a$ and $\bar{W}^b$.
Analogous to \cref{eqn:T,eqn:T-hat}, let
\begin{equation}
\label{eqn:T-BS}
\tilde{T}_f \equiv \GndBS f / \hat\sigma_f , \quad
\tilde{T}^{\mathcal{S}\vee} \equiv \sup_{f\in\mathcal{S}}\tilde{T}_f , \quad
\tilde{T}^{\mathcal{S}\wedge} \equiv \inf_{f\in\mathcal{S}}\tilde{T}_f , \quad
\abs{\tilde{T}}^{\mathcal{S}\vee} \equiv \sup_{f\in\mathcal{S}} \abs{\tilde{T}_f} .
\end{equation}
A bootstrap estimator of $\sigma_f$ suggested by \citet[p.\ 2222--2223]{ChernozhukovEtAl2013DR} is the scaled interquartile range.
Letting $z_\alpha$ denote the $\alpha$-quantile of the standard normal distribution and letting $(\GndBS f)_\alpha$ denote the $\alpha$-quantile of $\GndBS f$,
\begin{equation}
\label{eqn:sigma-hat-f-BS}
\hat\sigma_f = [(\GndBS f)_{0.75}-(\GndBS f)_{0.25}] / (z_{0.75}-z_{0.25}) .
\end{equation}

\Cref{res:BS} relies on Theorem 3.6.13 of \citet[p.\ 355]{vanderVaartWellner1996}.

\begin{theorem}
\label{res:BS}
Let the assumptions of \cref{res:sup} hold, as well as \Cref{a:weights} with the same $c$ for both samples.
Also, as in Theorem 3.6.13 of \citet[p.\ 355]{vanderVaartWellner1996}, assume measurability of the functions in $\mathcal{F}$ such that $\mathcal{F}_\delta$ is measurable for each $\delta>0$, where $\mathcal{F}_\delta \equiv \{ f-g : f,g\in\mathcal{F}, \rho_P(f-g)<\delta \}$ as in \citet[p.\ 350]{vanderVaartWellner1996} for both $P=P^a$ and $P=P^b$, with $\rho_P(f)\equiv\sqrt{P(f-Pf)^2}$ the variance seminorm as on page 89 of \citet{vanderVaartWellner1996}.
Then, conditional on almost all sequences of data, $\GndBS\weaklyto\Gd$ in $\ell^\infty(\mathcal{F})$.
Further,
\begin{equation*}
\tilde{T}^{\mathcal{S}\vee} \dconv T^{\mathcal{S}\vee} ,\quad
\tilde{T}^{\mathcal{S}\wedge} \dconv T^{\mathcal{S}\wedge},\quad
\abs{\tilde{T}}^{\mathcal{S}\vee} \dconv \abs{T}^{\mathcal{S}\vee} .
\end{equation*}
Consequently, the bootstrap critical values (quantiles) are consistent:
letting subscript $1-\alpha$ denote the $(1-\alpha)$-quantile of a random variable,
\begin{equation*}
\tilde{T}^{\mathcal{S}\vee}_{1-\alpha} 
\to %\pconv 
T^{\mathcal{S}\vee}_{1-\alpha} , \quad
\tilde{T}^{\mathcal{S}\wedge}_{1-\alpha} 
\to %\pconv 
T^{\mathcal{S}\wedge}_{1-\alpha} , \quad
\abs{\tilde{T}}^{\mathcal{S}\vee}_{1-\alpha} 
\to %\pconv 
\abs{T}^{\mathcal{S}\vee}_{1-\alpha} .
\end{equation*}
\end{theorem}
\begin{proof}
For $\GndBS$, the result is from Theorem 3.6.13 of \citet[p.\ 355]{vanderVaartWellner1996}, with the adjustment for $c$ in \cref{eqn:GndBS} above instead of in the limiting process.

For the suprema, the results hold by the continuous mapping theorem.

For the critical values (quantiles), convergence follows because the asymptotic distribution functions are all continuous and strictly increasing, so (for example) Lemma 21.2 of \citet{vanderVaart1998} applies.
\end{proof}

\Cref{sec:BScv} contains an example algorithm for computing bootstrap critical values.

\section{Additional Methods}
\label{sec:additional}

\Cref{sec:stepdown} describes a stepdown procedure to improve power of the basic MTP.
\Cref{sec:test} describes hypothesis tests of utility-based restricted stochastic dominance (and of non-dominance).
Proofs of theoretical results are in \cref{sec:app-pfs}.

\subsection{Stepdown Procedure}
\label{sec:stepdown}

A stepdown procedure in the spirit of \citet{Holm1979} \citep[see also][Ch.~9]{LehmannRomano2005text} can improve the basic MTP's power while maintaining strong control of FWER.
Although stepdown procedures are not uniformly better because in some cases FWER increases (but never above $\alpha$), they are often preferred.

\Cref{meth:stepdown} describes a stepdown procedure whose strong control of asymptotic FWER is given in \cref{res:stepdown}.
The general argument is the same as usual: if any initially-rejected $H_{0f}$ is true, then a familywise error is already committed, so further false rejections do not affect FWER; and if all initially-rejected $H_{0f}$ are false, then the critical value can be appropriately adjusted to reflect $\mathcal{F}_T$ being smaller than initially thought (because type I errors are only possible for $H_{0f}$ with $f\in\mathcal{F}_T$, and only the number of true $H_{0f}$ determines the necessary multiple testing adjustment).

\begin{method}[stepdown procedure]
\label{meth:stepdown}
The stepdown MTP proceeds as follows.
\begin{steps}
 \item\label{stepdown:0} Run 
\cref{meth:MTP-basic}: compute $\hat{T}_f$ as in \cref{eqn:T-hat} for all $f$, and reject all $H_{0f}$ for which $\hat{T}_f>\tilde{T}^{\mathcal{F}\vee}_{1-\alpha}$.
 Let $\hat{K}^{(0)}=\mathcal{F}$.  Let iteration counter $i=1$.
 \item\label{stepdown:1} Let $\hat{K}^{(i)}=\{f:H_{0f}\textrm{ not yet rejected}\}$.
 If $\hat{K}^{(i)}=\emptyset$ or $\hat{K}^{(i)}=\hat{K}^{(i-1)}$, then stop.
 \item\label{stepdown:recalibrate} Using the same bootstrap draws from \cref{stepdown:0}, use \cref{meth:BScv} with $\mathcal{S}=\hat{K}^{(i)}$ to compute bootstrap critical value $\tilde{T}^{\hat{K}^{(i)}\vee}_{1-\alpha}$.
 \item\label{stepdown:reject} Reject any additional $H_{0f}$ for which $\hat{T}_f > \tilde{T}^{\hat{K}^{(i)}\vee}_{1-\alpha}$.
 \item\label{stepdown:increment} Increment $i$ by one and return to \cref{stepdown:1}.
\end{steps}
\end{method}

\begin{proposition}
\label{res:stepdown}
\Cref{meth:stepdown} tests \cref{eqn:H0f} with strong control of asymptotic FWER at level $\alpha$.
\end{proposition}

\subsection{Testing Restricted Stochastic Dominance and Non-Dominance}
\label{sec:test}

\Cref{sec:test-SD} considers testing the null hypothesis of utility restricted stochastic dominance, whereas \cref{sec:test-nonSD} takes dominance as the alternative hypothesis (and thus non-dominance as the null).
As noted by \citet[pp.\ 88--89]{DavidsonDuclos2013}, strong and weak dominance cannot be distinguished statistically, so in this paper the choice of strict or weak inequality is made to optimize intuition or notational convenience.

Notationally, define pointwise $t$-statistics centered at zero (instead of at $P^af-P^bf$ as with $\hat{T}_f$ in \cref{eqn:T-hat}),
\begin{equation}
\label{eqn:T-hat-0}
\hat{T}^0_f \equiv
\sqrt{n_a}(\Pna-\Pnb)f/\hat\sigma_f .
\end{equation}
These are the $t$-statistics for testing $H_{0f}\colon P^af-P^bf=0$ (or $\le0$ or $\ge0$).

\subsubsection{Testing the null of utility restricted stochastic dominance}
\label{sec:test-SD}

Define the null hypothesis of stochastic dominance restricted to $\mathcal{F}$ (and the alternative of non-dominance) as
\begin{equation}
\label{eqn:H0-SDF}
\begin{split}
H_0&\colon Y^b \SDF Y^a
\iff
P^af-P^bf \le 0
\textrm{ for all }
f\in\mathcal{F} ,
\\
H_1&\colon Y^b \nonSDF Y^a
\iff
P^af-P^bf>0
\textrm{ for some }
f\in\mathcal{F} .
\end{split}
\end{equation}

\Cref{meth:test-SD-t} describes a test whose asymptotic size control is given in \cref{res:test-SD-t}.

\begin{method}[test of $\SDF$]
\label{meth:test-SD-t}
Compute $\hat{T}^0_f$ from \cref{eqn:T-hat-0} for each $f\in\mathcal{F}$.
In the notation of \cref{eqn:H0-SDF,eqn:T-BS}, reject $H_0\colon Y^b \SDF Y^a$ in favor of $H_1\colon Y^b \nonSDF Y^a$ when $\sup_{f\in\mathcal{F}} \hat{T}^0_f > \tilde{T}^{\mathcal{F}\vee}_{1-\alpha}$.
Equivalently: construct a one-sided $1-\alpha$ uniform confidence function $\hat{b}_1(\cdot)$ as in \cref{meth:band} and reject when $\hat{b}_1(f)>0$ for some $f\in\mathcal{F}$.
\end{method}

\begin{proposition}
\label{res:test-SD-t}
Given \cref{res:sup,res:BS}, \cref{meth:test-SD-t} has asymptotic size $\alpha$.
\end{proposition}

\subsubsection{Testing the null of non-dominance}
\label{sec:test-nonSD}

Now define the null and alternative hypotheses as
\begin{equation}
\label{eqn:H0-nonSDF}
\begin{split}
H_0&\colon Y^a \nonSDF Y^b
\iff
P^af-P^bf \le 0
\textrm{ for some }
f\in\mathcal{F} ,
\\
H_1&\colon P^af-P^bf>0
\textrm{ for all }
f\in\mathcal{F} .
\end{split}
\end{equation}

Testing \cref{eqn:H0-nonSDF} using the band in \cref{meth:band} is valid but conservative.
If $\hat{b}_1(f)>0$ for all $f\in\mathcal{F}$, then $\nonSDF$ can be rejected at level $\alpha$ in favor of dominance.
The band excludes the true $P^af-P^bf$ with probability $\alpha$, so given $P^af-P^bf<0$ for some $f$, there is only $\alpha$ probability of rejecting, i.e., of a band with $\hat{b}_1\ge0$.
The condition $\hat{b}_1(f)>0$ for all $f\in\mathcal{F}$ is equivalent to $\inf_{f\in\mathcal{F}} (\Pna-\Pnb)f-\tilde{T}^{\mathcal{F}\vee}_{1-\alpha} \hat\sigma_f/\sqrt{n_a}>0$, which is equivalent to $\inf_{f\in\mathcal{F}} \sqrt{n_a} (\Pna-\Pnb)f / \hat\sigma_f > \tilde{T}^{\mathcal{F}\vee}_{1-\alpha}$, or $\inf_{f\in\mathcal{F}} \hat{T}^0_f > \tilde{T}^{\mathcal{F}\vee}_{1-\alpha}$.

Power can be improved by decreasing the critical value $\tilde{T}^{\mathcal{F}\vee}_{1-\alpha}$ to the standard normal $z_{1-\alpha}$, while still controlling asymptotic size.
The idea is similar to Theorem 2.2 (pages 853--854) of \citet{KaurEtAl1994} for testing second-order restricted non-dominance over a range of CDF evaluation points (instead of utility functions).

\Cref{meth:test-nonSD-t} describes the test, whose asymptotic size control is given in \cref{res:test-nonSD-t}.

\begin{method}[test of $\nonSDF$]
\label{meth:test-nonSD-t}
Compute $\hat{T}^0_f$ from \cref{eqn:T-hat-0}, for $f\in\mathcal{F}$.
In the notation of \cref{eqn:H0-nonSDF}, reject $H_0\colon Y^a \nonSDF Y^b$ in favor of $H_1\colon Y^a \SDF Y^b$ when $\inf_{f\in\mathcal{F}} \hat{T}^0_f > z_{1-\alpha}$, where $z_{1-\alpha}$ is the $(1-\alpha)$-quantile of the $\NormDist(0,1)$ distribution.
\end{method}

\begin{proposition}
\label{res:test-nonSD-t}
Given \cref{res:sup}, the asymptotic size of the test in \cref{meth:test-nonSD-t} is $\alpha$.
\end{proposition}

\section{Proofs}
\label{sec:app-pfs}

\subsection{Proof of \texorpdfstring{\cref{res:Donsker-class-CRRA}}{Lemma \ref{res:Donsker-class-CRRA}}}

\begin{proof}
% VC: page 85ff (Sec 2.1) and Sec2.6 in vdvW96
First, the VC dimension is shown to be one by showing the subgraphs are ordered by inclusion.
This is most readily apparent by graphing\footnote{For example, \url{https://www.wolframalpha.com/input/?i=plot+\%28x\%5E\%281-0\%29-1\%29\%2F\%281-0\%29\%2C+\%28x\%5E\%281-0.5\%29-1\%29\%2F\%281-0.5\%29\%2C+\%28x\%5E\%281-2\%29-1\%29\%2F\%281-2\%29\%2C+\%28x\%5E\%281-4\%29-1\%29\%2F\%281-4\%29+from+x\%3D0..6}} but formally shown by the fact that $f_\theta(x)$ is (weakly) decreasing in $\theta$ for any $x$.
Taking the derivative with respect to $\theta$,
\begin{align}\notag
\pD{}{\theta}\frac{x^{1-\theta}-1}{1-\theta}
  &= \frac{(1-\theta)\pD{}{\theta}(x^{1-\theta}-1) - (x^{1-\theta}-1)(-1)}
          {(1-\theta)^2}
\\&= \frac{(\theta-1)x^{1-\theta}\ln(x) -1+x^{1-\theta}}%
          {(1-\theta)^2} .
\label{eqn:dCRRA-dtheta}
\end{align}
% R plot of derivative wrt theta:
% xx=1:444/100; th=6.6; plot(xx,-1+xx^(1-th)*(1+(th-1)*log(xx)), type='l', ylim=c(-5,1))
The numerator is now shown to be weakly negative (because the denominator is positive).
If $x=1$, then the numerator in \cref{eqn:dCRRA-dtheta} equals $(\theta-1)1^{1-\theta}\ln(1)-1+1^{1-\theta}=0$ for any $\theta$.
Further, the derivative with respect to $x$ of the numerator in \cref{eqn:dCRRA-dtheta} is negative for $x>1$ and positive for $x<1$:
\begin{align*}
\pD{}{x}
[(\theta-1)x^{1-\theta}\ln(x) -1+x^{1-\theta}]
  &= (\theta-1)[(1-\theta)x^{-\theta}\ln(x) + x^{1-\theta}(1/x)]
    +(1-\theta)x^{-\theta}
\\&= -(\theta-1)^2 x^{-\theta} \ln(x) 
    +(\theta-1)x^{-\theta}
    +(1-\theta)x^{-\theta}
\\&= [-(\theta-1)^2 \ln(x) 
      +(\theta-1)
      +(1-\theta)] x^{-\theta}
\\&= \overbrace{(\theta-1)^2}^{>0} [-\ln(x)] \overbrace{x^{-\theta}}^{>0}
\end{align*}
has the same sign as $-\ln(x)$, which is strictly positive for $0<x<1$ and strictly negative for $x>1$.
Thus, the zero value of \cref{eqn:dCRRA-dtheta} at $x=1$ is the global maximum for any $\theta$, so \cref{eqn:dCRRA-dtheta} is non-positive.
That is, $f_\theta(x)$ is decreasing in $\theta$ for any $x$.

Since the subgraphs are ordered by inclusion (because $f_a\le f_b$ for any $a\le b$), the VC dimension is one (VC index is two); e.g., see the proof of Lemma 2.6.16 in \citet{vanderVaartWellner1996}.
% rem: VC-index is smallest n for which no set of size n is shattered (\S2.6, p135, vdVW96) vs. VC-dimension is largest n for which some set of size n can be shattered (e.g., https://en.wikipedia.org/wiki/Vapnik%E2%80%93Chervonenkis_dimension#Definitions)

If additionally the envelope function is square integrable, then this implies $\mathcal{F}$ is $P$-Donsker by Theorems 2.5.2 and 2.6.7 in \citet{vanderVaartWellner1996}.
\end{proof}

\subsection{Proof of \texorpdfstring{\cref{res:MTP-basic}}{Proposition \ref{res:MTP-basic}}}

\begin{proof}
Using \cref{res:sup,res:BS},
\begin{align*}
\FWER &= \Pr( \textrm{reject any }H_{0f}
              \textrm{ with }
              f\in\mathcal{F}_T )
\\&=
\Pr( \sup_{f\in\mathcal{F}_T} \hat{T}_f 
    >\tilde{T}^{\mathcal{F}\vee}_{1-\alpha} )
\\&\le
\Pr( \sup_{f\in\mathcal{F}} \hat{T}_f 
    >\tilde{T}^{\mathcal{F}\vee}_{1-\alpha} )
\to \alpha .
\qedhere
\end{align*}
\end{proof}

\subsection{Proof of \texorpdfstring{\cref{res:stepdown}}{Proposition \ref{res:stepdown}}}

\begin{proof}
First, consider the infeasible oracle critical value from \cref{stepdown:recalibrate} when the true set of true hypotheses $\hat{K}^{(i)}=\mathcal{F}_T$ is used, i.e., the critical value $\tilde{T}^{\mathcal{F}_T\vee}_{1-\alpha}$.
Hypothetically, if this oracle critical value were used, then
\begin{equation}
\label{eqn:oracle-alpha}
\limn
\Pr( \overbrace{\sup_{f\in\mathcal{F}_T} \hat{T}_f}^{\hat{T}^{\mathcal{F}_T\vee}} > \tilde{T}^{\mathcal{F}_T\vee}_{1-\alpha} )
= \alpha ,
\end{equation}
because $\hat{T}^{\mathcal{F}_T\vee}\dconv T^{\mathcal{F}_T\vee}$ by \cref{res:sup} and $\tilde{T}^{\mathcal{F}_T\vee}_{1-\alpha} \pconv T^{\mathcal{F}_T\vee}_{1-\alpha}$, and the distribution of $T^{\mathcal{F}_T\vee}$ is continuous.

Second, the argument follows from a monotonicity property similar to (15.37) in \citet{LehmannRomano2005text}.
Specifically, $\mathcal{F}=\hat{K}^{(0)}\supset\hat{K}^{(1)}\supset\cdots$ because additional $H_{0f}$ can be rejected in every iteration, but once rejected they can never be un-rejected.
Combined with using the same bootstrap draws in every iteration, this implies the critical values are also monotonic over iterations: $\tilde{T}^{\hat{K}^{(0)}\vee}_{1-\alpha} > \tilde{T}^{\hat{K}^{(1)}\vee}_{1-\alpha} > \cdots$.

Consider a dataset in which
\begin{equation}
\label{eqn:oracle-nonreject}
\sup_{f\in\mathcal{F}_T}\hat{T}_f \le \tilde{T}^{\mathcal{F}_T\vee}_{1-\alpha}.
\end{equation}
That is, given the oracle critical value $\tilde{T}^{\mathcal{F}_T\vee}_{1-\alpha}$ based on the true $\mathcal{F}_T$, none of the true $H_{0f}$ is rejected (i.e., no familywise error is committed).
By \cref{eqn:oracle-alpha}, the probability of such a dataset converges to $1-\alpha$.

By induction, the stepdown procedure never commits a familywise error in such datasets.
Equivalently, it is shown that $\hat{K}^{(i)} \supseteq \mathcal{F}_T$ for all iterations $i$ in such datasets, which implies $\tilde{T}^{\hat{K}^{(i)}\vee}_{1-\alpha} \ge \tilde{T}^{\mathcal{F}_T\vee}_{1-\alpha}$.
In the first iteration of the stepdown procedure, $\hat{K}^{(0)}=\mathcal{F}\supseteq\mathcal{F}_T$, so the critical value satisfies $\tilde{T}^{\hat{K}^{(0)}\vee}_{1-\alpha} \ge \tilde{T}^{\mathcal{F}_T\vee}_{1-\alpha}$.
Next, it is shown that $\hat{K}^{(i)}\supseteq\mathcal{F}_T \implies \hat{K}^{(i+1)}\supseteq\mathcal{F}_T$.
Specifically, in iteration $i$, if $\hat{K}^{(i)}\supseteq\mathcal{F}_T$, then $\tilde{T}^{\hat{K}^{(i)}\vee}_{1-\alpha} \ge \tilde{T}^{\mathcal{F}_T\vee}_{1-\alpha}$, so none of the true $H_{0f}$ are rejected:
\begin{equation*}
\rlap{$\overbrace{\phantom{\sup_{f\in\mathcal{F}_T}
\hat{T}_f
\le \tilde{T}^{\mathcal{F}_T\vee}_{1-\alpha}}}^{\textrm{by \cref{eqn:oracle-nonreject}}}$}
\sup_{f\in\mathcal{F}_T}
\hat{T}_f
\le \underbrace{\tilde{T}^{\mathcal{F}_T\vee}_{1-\alpha}
\le \tilde{T}^{\hat{K}^{(i)}\vee}_{1-\alpha}}_{\textrm{since }\hat{K}^{(i)}\supseteq\mathcal{F}_T} .
\end{equation*}
Since no true $H_{0f}$ are rejected, $\hat{K}^{(i+1)}\supseteq\mathcal{F}_T$, too.
Thus,
\begin{equation*}
\FWER
= \Pr(\textrm{reject any true }H_{0f})
\le \overbrace{\Pr(\sup_{f\in\mathcal{F}_T}\hat{T}_f > \tilde{T}^{\mathcal{F}_T\vee}_{1-\alpha})}^{\to\alpha\textrm{ by \cref{eqn:oracle-alpha}}}
\to \alpha .
\qedhere
\end{equation*}
\end{proof}

\subsection{Proof of \texorpdfstring{\cref{res:CS}}{Proposition \ref{res:CS}}}

\begin{proof}
For the inner and outer CS, asymptotic coverage follows from FWER control.
For $\hat{\mathcal{C}}_1$, with $H_{0f}\colon P^af-P^bf\le0$,
\begin{align*}
\Pr(\hat{\mathcal{C}}_1\subseteq\mathcal{C})
  &= \Pr(f\in\hat{\mathcal{C}}_1\textrm{ only if }f\in\mathcal{C})
\\&= \Pr(\textrm{MTP rejects $H_{0f}$ only if }P^af-P^bf>0)
\\&= 1-\overbrace{\Pr(\textrm{MTP rejects any $H_{0f}$ with }P^af-P^bf\le0)}^{=\FWER\le\alpha+o(1)}
\\&\ge 1-\alpha+o(1).
\end{align*}

For $\hat{\mathcal{C}}_2$, with $H_{0f}\colon P^af-P^bf\ge0$,
\begin{align*}
\Pr(\hat{\mathcal{C}}_2\supseteq\mathcal{C})
  &= \Pr(f\in\hat{\mathcal{C}}_2\textrm{ when }f\in\mathcal{C})
\\&= \Pr(\textrm{MTP does not reject any $H_{0f}$ when }P^af-P^bf>0)
\\&\ge \Pr(\textrm{MTP does not reject any $H_{0f}$ when }P^af-P^bf\ge0)
\\&= 1-\overbrace{\Pr(\textrm{MTP rejects any $H_{0f}$ with }P^af-P^bf\ge0)}^{=\FWER\le\alpha+o(1)}
\\&\ge 1-\alpha+o(1).
\end{align*}

For the joint CS, coverage follows from the uniform confidence band's coverage:
\begin{align*}
\Pr(\hat{\mathcal{C}}_1\subseteq\mathcal{C}\subseteq\hat{\mathcal{C}}_2)
  &= \Pr\Bigl(\hat{b}_1(f)\le0\textrm{ whenever }P^af-P^bf\le0\textrm{, and } \\
  &\phantom{= \Pr\Bigl( \:\,}
    \hat{b}_2(f)>0\textrm{ whenever }P^af-P^bf>0 \Bigr)
\\&\ge \Pr\Bigl(\hat{b}_1(f)\le P^af-P^bf\textrm{ whenever }P^af-P^bf\le0\textrm{, and } \\
  &\phantom{= \Pr\Bigl( \:\,}
    \hat{b}_2(f)>P^af-P^bf\textrm{ whenever }P^af-P^bf>0 \Bigr)
\\&\ge \overbrace{\Pr(\hat{b}_1(f) \le P^af-P^bf\le\hat{b}_2(f)\textrm{ for all }f\in\mathcal{F})}^{\textrm{apply \cref{res:band}}}
\\&\ge 1-\alpha +o(1).
\qedhere
\end{align*}
\end{proof}

\subsection{Proof of \texorpdfstring{\cref{res:band}}{Proposition \ref{res:band}}}

\begin{proof}
For the symmetric band,
\begin{align*}
&
\Pr\{ \hat{b}_1(f) \le P^af-P^bf\le \hat{b}_2(f)\textrm{ for all }f\in\mathcal{F} \}
\\&=
\Pr\{ \overbrace{(\Pna-\Pnb)f 
- \abs{\tilde{T}}^{\mathcal{F}\vee}_{1-\alpha}
  \hat\sigma_f / \sqrt{n_a}}^{\hat{b}_1(f)\textrm{ from \cref{eqn:band-1s}}} 
  \le P^af-P^bf 
  \le \overbrace{(\Pna-\Pnb)f 
+ \abs{\tilde{T}}^{\mathcal{F}\vee}_{1-\alpha}
  \hat\sigma_f / \sqrt{n_a}}^{\hat{b}_2(f)\textrm{ from \cref{eqn:band-1s}}},
  \textrm{ }\forall f\in\mathcal{F} \}
\\&=
\Pr\{ 
- \abs{\tilde{T}}^{\mathcal{F}\vee}_{1-\alpha}
  \hat\sigma_f / \sqrt{n_a}
  \le [(P^a-P^b)-(\Pna-\Pnb)]f 
  \le \abs{\tilde{T}}^{\mathcal{F}\vee}_{1-\alpha}
  \hat\sigma_f / \sqrt{n_a}
  \textrm{ for all }f\in\mathcal{F} \}
\\&=
\Pr\{ 
- \abs{\tilde{T}}^{\mathcal{F}\vee}_{1-\alpha}
  \le \overbrace{\sqrt{n_a}[(P^a-P^b)-(\Pna-\Pnb)]f /\hat\sigma_f}^{-\hat{T}_f\textrm{ from \cref{eqn:T-hat}}}
  \le \abs{\tilde{T}}^{\mathcal{F}\vee}_{1-\alpha}
  \textrm{ for all }f\in\mathcal{F} \}
\\&=
\Pr\{ \abs{\hat{T}_f}
  \le \abs{\tilde{T}}^{\mathcal{F}\vee}_{1-\alpha}
  \textrm{ for all }f\in\mathcal{F} \}
\\&=
\Pr\{ \overbrace{\sup_{f\in\mathcal{F}}
     \abs{\hat{T}_f}}^{\abs{\hat{T}}^{\mathcal{F}\vee}\textrm{ from \cref{eqn:T-hat}}}
  \le \abs{\tilde{T}}^{\mathcal{F}\vee}_{1-\alpha} \}
\\&\to 1-\alpha
\end{align*}
because $\abs{\hat{T}}^{\mathcal{F}\vee} \dconv \abs{T}^{\mathcal{F}\vee}$ by \cref{res:sup} and $\abs{\tilde{T}}^{\mathcal{F}\vee}_{1-\alpha} \pconv \abs{T}^{\mathcal{F}\vee}_{1-\alpha}$ by \cref{res:BS}, and because $\abs{T}^{\mathcal{F}\vee}$ has a continuous distribution.

For the one-sided band with $\hat{b}_1$,
\begin{align*}
&
\Pr(\hat{b}_1(f) \le P^af-P^bf\textrm{ for all }f\in\mathcal{F})
\\&=
\Pr(\overbrace{(\Pna-\Pnb)f 
- \tilde{T}^{\mathcal{F}\vee}_{1-\alpha}
  \hat\sigma_f / \sqrt{n_a}}^{\hat{b}_1(f)\textrm{ from \cref{eqn:band-1s}}} - (P^af-P^bf) \le 0\textrm{ for all }f\in\mathcal{F})
\\&=
\Pr((\Pna-\Pnb)f - (P^af-P^bf) \le \tilde{T}^{\mathcal{F}\vee}_{1-\alpha}
  \hat\sigma_f / \sqrt{n_a}
  \textrm{ for all }f\in\mathcal{F})
\\&=
\Pr(\overbrace{\sqrt{n_a}[(\Pna-\Pnb)-(P^a-P^b)]f/\hat\sigma_f}^{\hat{T}_f\textrm{ from \cref{eqn:T-hat}}} \le \tilde{T}^{\mathcal{F}\vee}_{1-\alpha}
  \textrm{ for all }f\in\mathcal{F})
\\&=
\Pr(\overbrace{\sup_{f\in\mathcal{F}}
    \hat{T}_f}^{\hat{T}^{\mathcal{F}\vee}\textrm{ from \cref{eqn:T-hat}}} \le 
    \tilde{T}^{\mathcal{F}\vee}_{1-\alpha} )
\\&\to 1-\alpha
\end{align*}
because $\hat{T}^{\mathcal{F}\vee} \dconv T^{\mathcal{F}\vee}$ by \cref{res:sup} and $\tilde{T}^{\mathcal{F}\vee}_{1-\alpha} \pconv T^{\mathcal{F}\vee}_{1-\alpha}$ by \cref{res:BS}, and because $T^{\mathcal{F}\vee}$ has a continuous distribution.

For the one-sided band with $\hat{b}_2$,
\begin{align*}
&
\Pr(\hat{b}_2(f) \ge P^af-P^bf\textrm{ for all }f\in\mathcal{F})
\\&=
\Pr(\overbrace{(\Pna-\Pnb)f 
- \tilde{T}^{\mathcal{F}\wedge}_{\alpha}
  \hat\sigma_f / \sqrt{n_a}}^{\hat{b}_2(f)\textrm{ from \cref{eqn:band-1s}}} - (P^af-P^bf) \ge 0\textrm{ for all }f\in\mathcal{F})
\\&=
\Pr((\Pna-\Pnb)f - (P^af-P^bf) \ge \tilde{T}^{\mathcal{F}\wedge}_{\alpha}
  \hat\sigma_f / \sqrt{n_a}
  \textrm{ for all }f\in\mathcal{F})
\\&=
\Pr(\overbrace{\sqrt{n_a}[(\Pna-\Pnb)-(P^a-P^b)]f/\hat\sigma_f}^{\hat{T}_f\textrm{ from \cref{eqn:T-hat}}} \ge \tilde{T}^{\mathcal{F}\wedge}_{\alpha}
  \textrm{ for all }f\in\mathcal{F})
\\&=
\Pr(\overbrace{\inf_{f\in\mathcal{F}}
    \hat{T}_f}^{\hat{T}^{\mathcal{F}\wedge}\textrm{ from \cref{eqn:T-hat}}} \ge 
    \tilde{T}^{\mathcal{F}\wedge}_{\alpha} )
\\&\to 1-\alpha
\end{align*}
because $\hat{T}^{\mathcal{F}\wedge} \dconv T^{\mathcal{F}\wedge}$ by \cref{res:sup} and $\tilde{T}^{\mathcal{F}\wedge}_{1-\alpha} \pconv T^{\mathcal{F}\wedge}_{1-\alpha}$ by \cref{res:BS}, and because $T^{\mathcal{F}\wedge}$ has a continuous distribution.

For the ``equal-tailed'' band,
\begin{align*}
&\Pr(\hat{b}_1(f) \le P^af-P^bf \le \hat{b}_2(f)\textrm{ for all }f\in\mathcal{F})
\\&=
1-\Pr(\hat{b}_1(f)>P^af-P^bf\textrm{ for some $f$, or }\hat{b}_2(f)<P^af-P^bf\textrm{ for some }f)
\\&\ge
1 - \Pr(\hat{b}_1(f)>P^af-P^bf\textrm{ for some }f)
  - \Pr(\hat{b}_2(f)<P^af-P^bf\textrm{ for some }f)
\\&\to 1 - \alpha/2 - \alpha/2
= 1-\alpha ,
\end{align*}
essentially a Bonferroni adjustment argument.
\end{proof}

\subsection{Proof of \texorpdfstring{\cref{res:test-SD-t}}{Proposition \ref{res:test-SD-t}}}

\begin{proof}
Under $H_0$, $(P^a-P^b)f\le0$ for all $f\in\mathcal{F}$, so
\begin{equation*}
\hat{T}^0_f
= \frac{\sqrt{n_a}(\Pna-\Pnb)f}{\hat\sigma_f}
= \overbrace{\frac{\sqrt{n_a}[(\Pna-\Pnb)-(P^a-P^b)]f}{\hat\sigma_f}}^{\hat{T}_f}
 +\frac{\sqrt{n_a}\overbrace{(P^a-P^b)f}^{\le0}}{\hat\sigma_f}
\le 
\hat{T}_f
.
\end{equation*}
Thus, using notation from \cref{eqn:T,eqn:T-hat,eqn:T-BS} along with $\hat{T}^{\mathcal{F}\vee}\dconv T^{\mathcal{F}\vee}$ (\cref{res:sup}) and $\tilde{T}^{\mathcal{F}\vee}_{1-\alpha} \pconv T^{\mathcal{F}\vee}_{1-\alpha}$ (\cref{res:BS}), and the fact that $T^{\mathcal{F}\vee}$ has a continuous distribution,
\begin{align*}
\limn
\Pr(\sup_{f\in\mathcal{F}} \hat{T}^0_f > \tilde{T}^{\mathcal{F}\vee}_{1-\alpha})
&\le 
\limn
\Pr(\overbrace{\sup_{f\in\mathcal{F}} \hat{T}_f}^{\hat{T}^{\mathcal{F}\vee}} > \tilde{T}^{\mathcal{F}\vee}_{1-\alpha} )
\\&=
\limn
\Pr(\overbrace{\hat{T}^{\mathcal{F}\vee}}^{\dconv T^{\mathcal{F}\vee}} > \overbrace{\tilde{T}^{\mathcal{F}\vee}_{1-\alpha}}^{\pconv T^{\mathcal{F}\vee}_{1-\alpha}} )
\\&=
\Pr(T^{\mathcal{F}\vee} > T^{\mathcal{F}\vee}_{1-\alpha})
   =
\alpha .
\end{align*}

The version using $\hat{b}_1$ is equivalent because $\hat{b}_1(f)>0$ for some $f\in\mathcal{F}$ if and only if $\sup_{f\in\mathcal{F}} (\Pna-\Pnb)f-\tilde{T}^{\mathcal{F}\vee}_{1-\alpha}\hat\sigma_f/\sqrt{n_a}>0$, which in turn is equivalent to $\sup_{f\in\mathcal{F}} \sqrt{n_a}(\Pna-\Pnb)f/\hat\sigma_f>\tilde{T}^{\mathcal{F}\vee}_{1-\alpha}$ and thus $\sup_{f\in\mathcal{F}}\hat{T}^0_f>\tilde{T}^{\mathcal{F}\vee}_{1-\alpha}$.
\end{proof}

\subsection{Proof of \texorpdfstring{\cref{res:test-nonSD-t}}{Proposition \ref{res:test-nonSD-t}}}

\begin{proof}
Let $H_0$ hold, so $P^ag-P^bg\le0$ for at least one $g\in\mathcal{F}$.
Since $\{g\}\subset\mathcal{F}$, $\inf_{f\in\mathcal{F}}\hat{T}^0_f \le \hat{T}^0_g$.
Since $P^ag-P^bg\le0$,
\begin{equation*}
\hat{T}^0_g
= \frac{\sqrt{n_a}(\Pna-\Pnb)g}{\hat\sigma_g}
= \overbrace{\frac{\sqrt{n_a}[(\Pna-\Pnb)-(P^a-P^b)]g}{\hat\sigma_g}}^{\hat{T}_g}
 +\frac{\sqrt{n_a}\overbrace{(P^a-P^b)g}^{\le0}}{\hat\sigma_g}
\le 
\overbrace{\hat{T}_g \dconv \NormDist(0,1)}^{\textrm{by \cref{res:sup}}} .
\end{equation*}
Thus,
\begin{equation*}
\limn
\Pr(\inf_{f\in\mathcal{F}} \hat{T}^0_f > z_{1-\alpha})
\le 
\limn
\Pr(\hat{T}^0_g > z_{1-\alpha})
\le 
\limn
\Pr(\hat{T}_g > z_{1-\alpha})
=
\alpha .
\qedhere
\end{equation*}
\end{proof}

\section{Computing Bootstrap Critical Values}
\label{sec:BScv}

\Cref{meth:BScv} is an example algorithm to compute the bootstrap critical values from \cref{res:BS}.
Parts are similar to Algorithm 3 of \citet[p.\ 2222]{ChernozhukovEtAl2013DR}.
An example implementation in R \citep{R.core} is in the supplementary material and on my website.%
\footnote{\Blinded{[BLINDED]}{\url{https://kaplandm.github.io}}}

\begin{method}[bootstrap critical values]
\label{meth:BScv}
Take as given $\alpha\in(0,1)$, $\mathcal{S}\subseteq\mathcal{F}$, $n_a$, $n_b$, $c$ (from \cref{a:weights}), and the number of bootstrap replications $R$.
Note $c=1$ for empirical bootstrap, Bayesian bootstrap, and wild bootstrap with weights of unit variance.
In practice, usually $\mathcal{S}$ must be replaced by a finite grid, which may be arbitrarily fine (restricted only by computation time and patience).
\begin{steps}
 \item\label{BScv:wgt-a} Draw weights $(\tilde{W}^a_1,\ldots,\tilde{W}^a_{n_a})$.
 Compute $\PnaBS f=n_a^{-1}\sum_{i=1}^{n_a}f(Y^a_i)\tilde{W}^a_i$ for all $f\in\mathcal{S}$, as well as $\bar{W}^a=n_a^{-1}\sum_{i=1}^{n_a}\tilde{W}^a_i$.
 \item\label{BScv:wgt-b} Independently draw $(\tilde{W}^b_1,\ldots,\tilde{W}^b_{n_b})$.
 Compute $\PnbBS f=n_b^{-1}\sum_{i=1}^{n_b}f(Y^b_i)\tilde{W}^b_i$ for all $f\in\mathcal{S}$, as well as $\bar{W}^b=n_b^{-1}\sum_{i=1}^{n_b}\tilde{W}^b_i$.
 \item\label{BScv:GndBS} Following \cref{eqn:GndBS}, compute $\GndBS f=\sqrt{n_a}[(\PnaBS -\PnbBS )-(\bar{W}^a\Pna -\bar{W}^b\Pnb )]f / c$ for all $f\in\mathcal{S}$.
 \item Repeat \cref{BScv:wgt-a,BScv:wgt-b,BScv:GndBS} $R$ times, generating $(\GndBS f)^{(r)}$ in bootstrap replications $r=1,\ldots,R$.
 \item\label{BScv:sigma-hat} For all $f\in\mathcal{S}$, compute $\hat\sigma_f$ following \cref{eqn:sigma-hat-f-BS} with quantiles from $((\GndBS f)^{(1)},\ldots,(\GndBS f)^{(R)})$.
 \item For all $f\in\mathcal{S}$ and $r=1,\ldots,R$, compute $\tilde{T}^{(r)}_f=(\GndBS f)^{(r)} / \hat\sigma_f$ as in \cref{eqn:T-BS}.
 \item Following \cref{eqn:T-BS}, for each $r=1,\ldots,R$, compute
 \[ \tilde{T}^{\mathcal{S}\vee}_{(r)} = \sup_{f\in\mathcal{S}} \tilde{T}^{(r)}_f , \quad
 \tilde{T}^{\mathcal{S}\wedge}_{(r)} = \inf_{f\in\mathcal{S}} \tilde{T}^{(r)}_f , \quad
 \abs{\tilde{T}}^{\mathcal{S}\vee}_{(r)} = \sup_{f\in\mathcal{S}} \abs{\tilde{T}}^{(r)}_f . \]
 \item Let $\tilde{T}^{\mathcal{S}\vee}_{1-\alpha}$ be the sample $(1-\alpha)$-quantile of $(\tilde{T}^{\mathcal{S}\vee}_{(1)},\ldots,\tilde{T}^{\mathcal{S}\vee}_{(R)})$, 
 $\tilde{T}^{\mathcal{S}\wedge}_{1-\alpha}$ the sample $(1-\alpha)$-quantile of $(\tilde{T}^{\mathcal{S}\wedge}_{(1)},\ldots,\tilde{T}^{\mathcal{S}\wedge}_{(R)})$, and
 $\abs{\tilde{T}}^{\mathcal{S}\vee}_{1-\alpha}$ the sample $(1-\alpha)$-quantile of $(\abs{\tilde{T}}^{\mathcal{S}\vee}_{(1)},\ldots,\abs{\tilde{T}}^{\mathcal{S}\vee}_{(R)})$.
\end{steps}
\end{method}

\section{Additional Empirical Examples}
\label{sec:app-emp}

This section includes additional empirical examples.

\subsection{Job Training: JTPA}
\label{sec:emp-JTPA}

I use the same sample of Job Training Partnership Act (JTPA) data analyzed by \citet{AbadieEtAl2002} and \citet{KaplanSun2017}, who both find larger effects at larger quantiles.
The general setup, interpretation, and caveats are the same as for the NSW job training example in \cref{sec:emp-NSW}.

\begin{figure}[htbp]
\centering
\hfill
\includegraphics{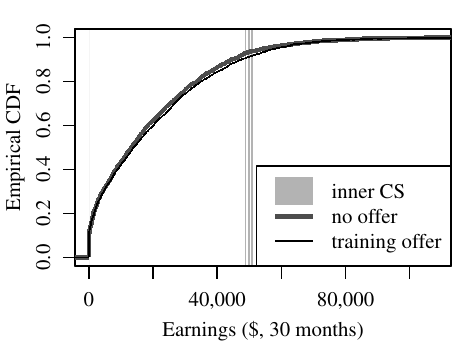}
\hfill
\includegraphics{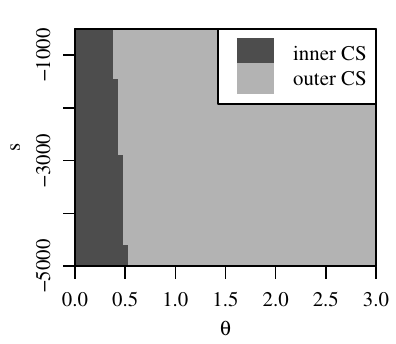}
\hfill\null
\caption{Earnings by job training offer status.}
\label{fig:ex-JTPA}
\end{figure}

\Cref{fig:ex-JTPA} shows the JTPA results.
The outcome variable is 30-month earnings after the training period.
Because the intention-to-treat effects are relatively small, $\alpha=0.1$ is used.
The left graph shows that the training-assigned empirical CDF is below the not-assigned empirical CDF everywhere, but by a slim margin.
Consequently, the CDF-based inner CS is very small, containing only values near \USD{0} and \USD{50000} (around the $90$th percentile).

\Cref{fig:ex-JTPA}'s right graph shows the utility function (social welfare function) inner CS for the JTPA data.
Here, the shift $s$ includes negative values to account for the fact that even individuals with zero earnings do not have zero consumption.
Only small values of $\theta$ (below $0.5$) are included in the inner CS.
This makes sense because the lower part of the training-assigned distribution is actually worse: although it has slightly fewer zero values ($10.3\%$ instead of $10.4\%$), it has lower $11$th through $18$th percentile values than the not-assigned distribution.
(The difference is not large enough to make the opposite conclusion, so all points in the graph are still inside the outer CS.)
Although possibly this is due to sampling error, it makes it impossible to be statistically certain that the training-assigned distribution is better with any substantial degree of inequality aversion.
But, due to having larger upper percentiles, the training distribution is preferred at a relatively high confidence level given social welfare functions closer to so-called ``utilitarian'' social welfare with $\theta=0$.
Altogether, there is strong evidence that the training-assigned distribution is better if we define ``better'' using a relatively small $\theta$ (from the inner CS), but not strong evidence of a broad consensus, although there is also no strong evidence \emph{against} a broad consensus given the large outer CS.

\subsection{Union Membership}
\label{sec:emp-union}

\begin{figure}[htbp]
\centering
\hfill
\includegraphics{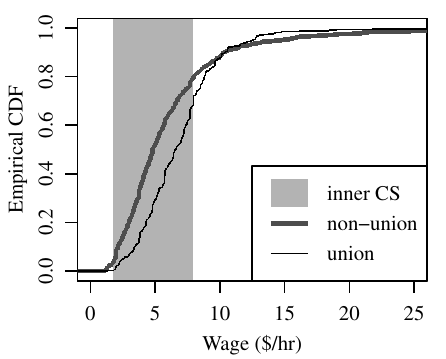}
\hfill
\includegraphics{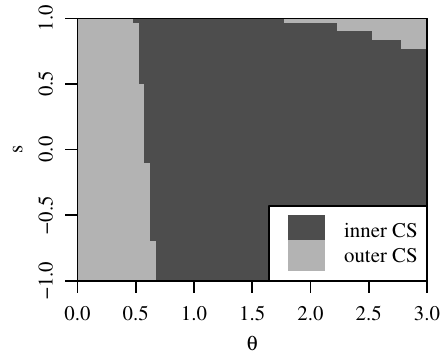}
\hfill\null
\caption{Wage by union membership.}
% $\alpha=0.005$
\label{fig:ex-beauty-union}
\end{figure}

\Cref{fig:ex-beauty-union} shows the results for hourly wage distributions by union membership using the \code{wooldridge} package's dataset \code{beauty} \citep[originally from][]{HamermeshBiddle1994}, with $\alpha=0.005$ due to the large sample size ($n=1260$).
Again, the question is not whether a specific existing individual should join a union, but rather the social question of whether we prefer the union or non-union wage distribution, using the thought experiment of taking a random draw from one distribution or the other (in the spirit of Vickrey, Harsanyi, and Rawls; see \cref{ft:OP}), or equivalently the Atkinson social welfare function.
The union wage empirical CDF lies below the non-union empirical CDF over most of the distribution (by a statistically significant amount, as shown by the CDF-based inner CS), but the empirical CDFs cross around the $90$th percentile.
This pattern is reflected by the large inner CS in the right graph of \cref{fig:ex-beauty-union} that in particular contains larger values of $\theta$ (more inequality-averse).
This shows a broad consensus preference for the union wage distribution among all but the least inequality-averse social welfare functions.
Further, with larger $\alpha=0.05$, the utility functions with $\theta\le0.5$ are also included in the inner CS.

\end{document}